\documentclass[smallextended]{svjour3} 
\usepackage[utf8]{inputenc}
\usepackage{pgf,tikz,pgfplots}
\usetikzlibrary{calc}
\pgfplotsset{compat=1.15}
\usetikzlibrary{arrows}

\usepackage{enumerate}
\usepackage{pifont}
\usepackage{grffile}
\usepackage[normalem]{ulem}
\usepackage{amsfonts}
\usepackage{enumitem}
 \usepackage{setspace}
 \expandafter\let\csname equation*\endcsname\relax
\expandafter\let\csname endequation*\endcsname\relax
\usepackage[utf8]{inputenc}
\usepackage{BOONDOX-cal}
\usepackage{bbm}
\usepackage{epsfig}
\usepackage{color}
\usepackage{bm}
\usepackage{float}
\usepackage{physics}
\usepackage{bigints}
\usepackage{amssymb}
\usepackage{subfigure}
\usepackage{setspace}
\usepackage{textcomp}
\usepackage{amsmath}
\usepackage{esint}
\usepackage{xcolor}
\usepackage{subfigure}
\usepackage{float}
\usepackage{soul}
\usepackage{stackrel}
\usepackage{color}
\usepackage[bottom]{footmisc}
\usepackage[margin=1in]{geometry}
\newcommand{\zbar}{\raisebox{0.2ex}{--}\kern-0.6em Z}
\usepackage[
colorlinks=true, 
pdfstartview=FitV, 
linkcolor=blue, 
citecolor=magenta, 
urlcolor=magenta, 
bookmarks=true,
bookmarksnumbered=true,
pdftitle={},
pdfauthor={}
]{hyperref}
\usepackage{epsfig,amsopn}
\usepackage{graphicx}

\definecolor{pink}{rgb}{1,0,0.6}
\definecolor{dgreen}{rgb}{0,0.7,0}

\usepackage{epsfig,amsopn}
\usepackage{graphicx}
\usepackage{titlesec}
\titleformat{\subsection}{\bfseries}{\thesubsection}{1em}{}

\begin{document}

\title{Generalized hydrodynamics and approach to Generalized Gibbs equilibrium for a classical harmonic chain}

\author{Saurav Pandey\textsuperscript{1}, Abhishek Dhar\textsuperscript{1} and Anupam Kundu\textsuperscript{1}}

\institute{${}^{1}$ International Centre for Theoretical Sciences, Tata Institute of Fundamental Research, Bangalore 560089, India}

\date{\today}
\authorrunning{GHD and approach to GGE for a classical harmonic chain}
\maketitle

\begin{abstract}
We study the evolution of a classical harmonic chain with nearest-neighbor interactions starting from domain wall initial conditions. The initial state is taken to be either a product of two Gibbs Ensembles (GEs) with unequal temperatures on the two halves of the chain or a product of two Generalized Gibbs Ensembles (GGEs) with different parameters in the two halves. For this system, we construct the Wigner function and demonstrate that its evolution defines the Generalized Hydrodynamics (GHD) describing the evolution of the conserved quantities. We solve the GHD for both finite and infinite chains and compute the evolution of conserved densities and currents. For a finite chain with fixed boundaries, we show that these quantities relax as $\sim 1/\sqrt{t}$ to their respective steady-state values given by the final expected GE or GGE state, depending on the initial conditions. Exact expressions for the Lagrange multipliers of the final expected GGE state are obtained in terms of the steady state densities. 
In the case of an infinite chain, we find that the conserved densities and currents at any finite time exhibit ballistic scaling while, at infinite time,  any finite segment of the system can be described by a current-carrying non-equilibrium steady state (NESS). We compute the scaling functions analytically and show that the relaxation to the NESS occurs as $\sim 1/t$ for the densities and as $\sim 1/t^2$ for the currents. We compare the analytic results from hydrodynamics with those from exact microscopic numerics and find excellent agreement.
\end{abstract}


\tableofcontents

\section{Introduction} \label{sec:intro}

Thermalization or the absence of it in isolated Hamiltonian systems has been an active area of research, ever since the first numerical study of the FPUT model \cite{fput1955,dauxois2008} was carried out. For generic non-integrable systems, energy is typically the only conserved quantity. Such a system usually thermalizes to a Gibbs Ensemble (GE) in the sense that, starting from a non-equilibrium initial state, the macroscopic properties of the system relax to the values determined by the GE. On the other hand, integrable systems have an extensive number of conserved quantities and thus do not thermalize to a GE. However, they are expected to relax to a steady state described by a Generalized Gibbs Ensemble (GGE). The relaxation and approach to a steady state has been shown in non-interacting integrable models such as the hard particle gas (HPG) \cite{chakraborti2022} and the harmonic chain \cite{vulpiani2022} and in interacting integrable models such as the alternate mass hard particle gas (AMHPG) \cite{chakraborti2023} and the Toda chain \cite{ganapa2020}. Thermalization for a low-density free fermion lattice model is shown rigorously in \cite{tasaki2023} while relaxation of ideal quantum gases is studied in \cite{pandey2023}. The approach to GGE however has been demonstrated mainly in quantum integrable models \cite{rigol2007,rigol2011,calabrese2011,essler2022,pozsgay2014quantum,pozsgay2016real} and recently for the classical hard rods gas \cite{sahil2023}. 

An effective understanding of a system's macroscopic evolution and relaxation to equilibrium can be achieved via the hydrodynamic approach. The hydrodynamic theory is based on identifying the conserved quantities for writing down the continuity equations involving densities and currents and the assumption of local equilibrium for expressing currents in terms of the densities. For non-integrable systems, one writes the hydrodynamic equations for the few conserved quantities such as mass, momentum, and energy. In contrast, integrable systems with an extensive number of conservation laws are described by Generalized Hydrodynamics (GHD)  which is written in terms of the quasi-particle density. One needs to distinguish between non-interacting and interacting integrable systems. For the former case, the hydrodynamic equations are easier to write since the quasi-particles do not undergo phase shifts upon collisions. Common examples include the harmonic chain for which hydrodynamic equations were derived both at the Euler level~\cite{dobrushin1986} (see also \cite{mielke2006}) and with higher derivative  corrections~\cite{dobrushin1988}, and the HPG \cite{chakraborti2022}.  On the other hand for the interacting case, quasi-particles undergo phase shifts upon collision. The simplest example of such systems is a collection of hard rods for which the hydrodynamic equations were written long back~\cite{percus1969exact,boldrighini1983euler,boldrighini1997navierstokes}. For the case of interacting integrable quantum systems, GHD was derived more recently~\cite{castro2016emergent,bertini2016transport} and quite remarkably they have the same structure as the classical systems~\cite{doyon2020}. A number of studies have established GHD  for several interacting classical and quantum integrable systems such as Toda chain~\cite{spohn2021toda,cao2019gge,mazzuca2023equilibrium,kundu2023integrable}, the $\delta$-Bose gas~\cite{ruggiero2020} and hard rods \cite{doyon2017,sahil2023}. GHD has been successfully used to understand unusual equilibration of trapped integrable systems~\cite{caux2019hydrodynamics,schemmer2019generalized,cao2018incomplete} and also the predictions of GHD have been experimentally verified \cite{malvania2021generalized}. 

The harmonic chain provides us with an example of an integrable system that is analytically tractable and allows macroscopic description in terms of hydrodynamics \cite{dobrushin1986,dobrushin1988}. For this system, a detailed comparison of the microscopic and hydrodynamic calculations is possible. Approach to GGE starting from a domain wall initial condition with unequal temperatures in the two halves has been studied for the case of an infinite quantum harmonic chain in \cite{eisler2014} where the Lagrange multipliers of the final GGE state were evaluated explicitly. Convergence of the phase space distribution to a Gaussian stationary measure, starting from an initial state which is also Gaussian, is studied in \cite{dudnikova2003convergence}. In \cite{boldrighini1983conservedQ}, two families of conserved quantities for infinite harmonic lattices were shown and the convergence of the covariance matrix to its stationary form was studied. Hydrodynamic approaches have also been applied to harmonic systems. In \cite{dobrushin1986}, starting from a family of initial states that are slowly varying in space, the hydrodynamic equation of an infinite classical harmonic chain was derived at the Euler level in terms of what is referred to as the spectral density matrix function (SDMF) in the paper, which is essentially the Fourier transform of the correlation matrix defined in a coarse-grained cell around a macroscopic point. The study was extended in \cite{dobrushin1988} where the next-order correction (in space-time scaling parameter) to the Euler equation was derived. Alternatively, macroscopic evolution in the harmonic chain has been studied using Wigner functions \cite{spohn2006boltzmann,komorowski2020high,dudnikova2005local,bernardin2019}. The harmonic chain being an integrable system, one expects a generalized hydrodynamic description for the macro-evolution. However, the connections between the correlation matrix approach \cite{dobrushin1986,dobrushin1988} and the Wigner function method \cite{spohn2006boltzmann} have not been elaborated in the physics literature. Furthermore, and somewhat surprisingly, their relationship to GHD remains unexplored. In this paper, we aim to address this gap and elucidate these connections which we use to understand non-equilibrium evolution and approach to the stationary state. 

In this paper, we consider a classical harmonic chain with nearest-neighbor interactions either of finite length (with fixed boundary conditions) or of infinite extent. Initially, we prepare the system in a domain wall configuration by taking the two halves of the chain either in thermal equilibrium with unequal temperatures or in different GGEs with unequal parameters (Lagrange multipliers). Starting from this initial condition, we study the evolution of the chain and the approach to the stationary state in the long time limit (for the finite case). 
Following is a summary of our main results: 
\begin{itemize}[label=$\bullet$]
\item We clarify how the Wigner function method \cite{spohn2006boltzmann} is related to the correlation matrix approach in \cite{dobrushin1986}. In particular, we write an explicit expression for the Wigner function in terms of the Fourier transform of the local correlation function --- this allows us to show the equivalence between the Euler equation in \cite{dobrushin1986} and the equation satisfied by the Wigner function [see Sec.~(\ref{subsec:wignerCorrMat})].
\item We express all the local conserved densities and currents in terms of the Wigner function which allows us to write the hydrodynamic equations for all the conserved quantities. The equation for the coarse-grained Wigner function can thus be considered as the GHD equation for the harmonic chain. This approach is similar to that used for deriving the GHD equation for free fermions~\cite{essler2022}.

\item Using the non-interacting structure, we solve the GHD equations analytically and obtain explicit expressions for the time evolution of the density and current profiles starting from non-equilibrium initial conditions. We compare these solutions with microscopic computations and find remarkable agreement.

\item On the infinite line, the domain wall spreads ballistically and from our solution of GHD equations we can compute the exact scaling functions corresponding to both the density and current profiles. At any fixed location, the density and the current corresponding to any conserved quantity approach a stationary finite value in the $t \to \infty$ limit. The approach to the steady state value is $\sim 1/t$ and $\sim 1/t^2$, for densities and currents respectively. Therefore the stationary state inside any finite segment should be describable by a GGE which, in addition to the conserved densities, also includes all the associated currents that are conserved in the infinite geometry \cite{eisler2014}.

\item For the domain wall initial condition on a finite chain, we observe that the chain relaxes to a final stationary state that is consistent with a GGE state in which all the conserved densities become homogeneous in space and stationary in time while the corresponding currents vanish. Note that in this case, current conservation is violated by the reflections at the boundaries. The conserved densities and the associated currents approach their stationary values as $1/\sqrt{t}$ in the $t \to \infty $ limit. 

\item The final expected GGE state in the finite chain is characterized by chemical potentials $\{\lambda_n\}$ that are completely determined by the values of the conserved quantities. We provide explicit expressions of these chemical potentials in terms of the conserved densities. For the particular case of the domain wall initial condition composed of two Gibbs states at different temperatures, the final stationary state is also a Gibbs state. For the other case studied in this work in which the initial domain wall state is composed of two GGEs where only the first two Lagrange parameters are non-zero in either half of the chain, we find that the final state is also expected to be a GGE with Lagrange parameters that decay exponentially in strength for large $n$. 
\end{itemize}

The rest of the paper is organized as follows. In Sec.~\eqref{sec:microDynamics}, we introduce our system of a classical harmonic chain with nearest-neighbor interactions and study its microscopic evolution averaged over initial states drawn from a distribution that is a product of two Generalized Gibbs Ensembles with different values for the chemical potentials in the two halves of the chain. The Wigner function for harmonic lattices is introduced in Sec.~\eqref{subsec:wignerCorrMat} and its connection to the correlation matrix approach used in the literature is made explicit. In Sec.~\eqref{subsec:conservedGHD}, we construct the conserved quantities for the harmonic chain and express the corresponding densities and currents in terms of the Wigner function. In Sec.~\eqref{sec:domainWall}, we derive the hydrodynamic evolution for the conserved quantities using the Wigner function approach and compare them with the exact microscopic results. The case where the two halves are initially described by a GE with different temperatures is studied in Sec.~\eqref{subsec:domainWallGE} whereas Sec.~\eqref{subsec:domainWallGGE} deals with the case where the two halves are initially described by a GGE with only the first two parameters being nonzero and unequal in the left and right parts of the chain. We conclude our findings in Sec.~\eqref{sec:conclusions}.

\section{Microscopic description of the system} \label{sec:microDynamics}
\begin{figure}[H]
\begin{center}
\begin{tikzpicture}[>=latex]
  \tikzset{
    particlered/.style={circle, draw, fill=red, inner sep=1pt, minimum size=10pt},
    particleblue/.style={circle, draw, fill=blue, inner sep=1pt, minimum size=10pt},
    spring/.style={thick, decorate, decoration={zigzag, segment length=6, amplitude=2}}
  }

  \foreach \i in {1,...,4}
    \node[particlered] (particle\i) at (\i,0) {};
  \foreach \i in {5,...,8}
    \node[particleblue] (particle\i) at (\i,0) {};

  \foreach \i in {1,3,4,5,7} {
    \draw[spring][line width=1pt] (particle\i) -- (particle\the\numexpr\i+1\relax);
    }
    
  \draw[spring][line width=1pt] (0,0) -- (particle1);
  
  \draw[spring][line width=1pt] (9,0) -- (particle8);

  \draw[line width=2pt] (0,-0.5) -- (0,0.5);

  \draw[line width=2pt] (9,-0.5) -- (9,0.5);

  \draw[line width=2pt] (4.5,-0.5) -- (4.5,0.5);

  \foreach \i in {1,...,8}
    \filldraw (\i,0) circle (0pt);

  \node at (2.5,0) {.....};
  
  \node at (6.5,0) {.....};

  \node[above, font=\large] at (2.5,0.35) {$\{ \lambda_n^L \}$};
  \node[above, font=\large] at (6.5,0.35) {$\{ \lambda_n^R \}$};
  \node[right, font=\large] at (9.5,0) {$t=0$};
  \foreach \i in {1,2}
    \node[below] at (\i,-0.35) {\fontsize{10}{12}\selectfont $\i$};
  \node[below] at (3,-0.35) {\fontsize{10}{12}\selectfont N$-$1};
  \node[below] at (4,-0.35) {\fontsize{10}{12}\selectfont N};
  \node[below] at (5,-0.35) {\fontsize{10}{12}\selectfont N$+$1};
  \node[below] at (6,-0.35) {\fontsize{10}{12}\selectfont N$+$2};
  \node[below] at (7,-0.35) {\fontsize{10}{12}\selectfont 2N$-$1};
  \node[below] at (8,-0.35) {\fontsize{10}{12}\selectfont 2N};
\end{tikzpicture}
\end{center}

\begin{center}
\begin{tikzpicture}[>=latex]
  \foreach \i in {1,...,8} {
    \pgfmathsetmacro\red{1-\i/8}
    \pgfmathsetmacro\blue{(\i-1)/8}
    \definecolor{mycolor}{rgb}{\red,0.3,\blue}
    \node[circle, draw, fill=mycolor, inner sep=0pt, minimum size=10pt,     
        postaction={
        fill=white,
        opacity=0.2,
      }] (particle\i) at (\i,0) {};
    }
    \tikzset{spring/.style={thick, decorate, decoration={zigzag, segment length=6, amplitude=2}}
  }
  \foreach \i in {1,3,4,5,7}
    \draw[spring][line width=1pt] (particle\i) -- (particle\the\numexpr\i+1\relax);

    \draw[spring][line width=1pt] (0,0) -- (particle1);
  
    \draw[spring][line width=1pt] (9,0) -- (particle8);

  \draw[line width=2pt] (0,-0.5) -- (0,0.5);

  \draw[line width=2pt] (9,-0.5) -- (9,0.5);

  \foreach \i in {1,...,8}
    \filldraw (\i,0) circle (0pt);

  \node at (2.5,0) {.....};
  
  \node at (6.5,0) {.....};

    \node[right, font=\large] at (9.5,0) {$t>0$};
\end{tikzpicture}
\end{center}
\caption{Plot showing a schematic of the harmonic chain system of length $2N$. Initially, at $t=0$, the left and right halves are disconnected and in general, described by two GGEs with unequal chemical potentials $\lambda_n^L \neq \lambda_n^R$. At $t>0$, the two halves interact and the system evolves as a whole towards a new stationary state.} \label{schematic}
\end{figure}
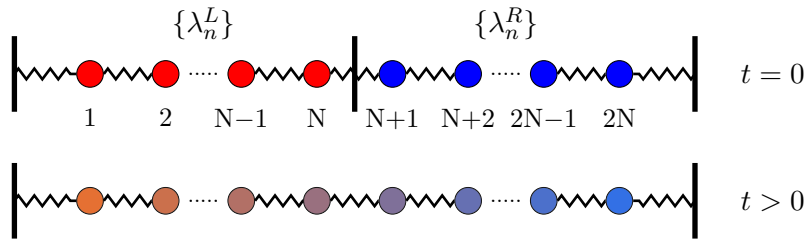

In this section, we define the microscopic model of our classical harmonic chain and state the exact expressions for various correlation functions that can be obtained from the solution of the microscopic dynamics. We consider a chain of $2N$ particles of unit masses whose positions and momenta are denoted by $\{ q_j, p_j \}, ~j=1,\ldots 2N$.  Assuming fixed boundary conditions: $q_0 = q_{2N+1} = 0$, the Hamiltonian is given by
\begin{equation}
H = \sum\limits_{j=1}^{2N} \frac{p_j^2}{2} + \sum\limits_{j=0}^{2N} \frac{(q_{j+1} - q_j)^2}{2} = \displaystyle{\frac{1}{2}} \left( \mathbf{p}^T \mathbf{p} + \mathbf{q}^T \Phi \mathbf{q} \right), \label{hamiltonian}
\end{equation}
where $\mathbf{q} = \{ q_1, \ldots, q_{2N} \}$ and $\mathbf{p} = \{ p_1, \ldots, p_{2N} \}$ are vectors containing the particle positions and momenta respectively and $\Phi$ denotes the force matrix with elements $\Phi_{j, j'} = 2 \delta_{j, j'} - \delta_{j+1, j'} - \delta_{j-1, j'}$ for $j,j'=1,2,...,2N$, where $\delta_{jj'}$ is Kronecker delta. This gives us the equations of motion (EOM) 
\begin{align}
\ddot{\mathbf{q}} &= -\Phi \mathbf{q}. \label{eom}
\end{align} 
Let $U$ be the matrix of eigenfunctions that diagonalises $\Phi$ {\it i.e.,} $U^T \Phi U = \Omega^2$ where $\Omega^2_{k, k'} = \omega_k^2 \delta_{k, k'}$. The eigenvalues and eigenfunctions are given by 
\begin{subequations}
\begin{align}
&\omega_{k_\ell}^2 = 2(1 - \cos k_{\ell}), \label{w_k}\\
&\psi_{k_\ell}(j) = U_{jk_\ell} = \displaystyle{\sqrt{\frac{2}{2N+1}}} \sin j k_{\ell}, \label{psi_k} \\
&\text{where}~k_\ell = \displaystyle{\frac{\pi \ell}{2N+1}}, ~\ell = 1, \ldots, 2N. \label{k_ell}
\end{align}
\end{subequations}
We can write down a general solution to Eq.~\eqref{eom} in matrix notation as 
\begin{subequations}
\begin{align}
\mathbf{q}(t) &= \dot{A}(t) \mathbf{q}(0) +  A(t) \mathbf{p}(0), \\
\mathbf{p}(t) &=  \ddot{A}(t) \mathbf{q}(0) + \dot{A}(t) \mathbf{p}(0),
\end{align}
\label{q_t-p_t}
\end{subequations}
where we have defined
\begin{align}
A(t) = U \displaystyle{\frac{\sin (\Omega t)}{\Omega}} U^T.
\label{A(t)}
\end{align}
\noindent 
For a harmonic chain of $N$ particles, one can construct $N$ number of local conserved quantities which we denote by $Q^{(n)}({\bf p,q}),~n=0,1,2,...,N-1$. These conserved quantities are of the form: 
\begin{align}
Q^{(n)}({\bf p,q}) = \frac{1}{2} \left( \mathbf{p}^T B^{(n)} \mathbf{p} + \mathbf{q}^T M^{(n)} \mathbf{q} \right),
\label{def:Q^n}
\end{align} 
\noindent
where $B^{(n)}$ and $M^{(n)}$ are $N \times N$ symmetric matrices [see Sec.~\eqref{subsec:conservedGHD} for a detailed discussion]. The general stationary state of the chain is therefore described by a GGE of the form
\begin{align}
P_{\rm GGE}({\bf{p,q}}) = \frac{1}{Z(\{ \lambda_n \})} \exp \left( -\sum\limits_{n=0}^{N-1} \lambda_n Q^{(n)} \right), \label{P_GGE}
\end{align}
where $\lambda_n$ are the corresponding Lagrange multipliers that determine the average values of the conserved quantities. Here, $Z(\{ \lambda_n \})$ is the GGE partition function for a chain of $N$ particles, which is explicitly given by [see Appendix~\eqref{app:Z_GGE}]
\begin{align}
Z(\{ \lambda_n \})=(2\pi)^N \prod_{\ell=1}^N \frac{1}{\omega_{k_{\ell}}\sum_{n=0}^{N-1} \lambda_{n}\cos\left(n k_{\ell}\right)},
\label{partitionFunc}
\end{align}
where  $\omega_k$ and $k_\ell$  are given in Eqs.~(\ref{w_k},~\ref{k_ell}), respectively.\\

\noindent
\textbf{Initial condition:}
As shown in the schematic in Fig.~\eqref{schematic}, our system comprises of two such $N$-particle chains that are initially in a domain wall configuration described by a product of two GGE states 
\begin{align}
P_{\rm in} = \frac{e^{-\sum_n \lambda_n^L Q^{(n)}_L}}{Z^L} \times \frac{e^{-\sum_n \lambda_n^R Q^{(n)}_R}}{Z^R}, \label{icLR}
\end{align}
with different values of the Lagrange multipliers in the left and right halves, denoted respectively by $\{\lambda^{L}_n\}$ and $\{\lambda^{R}_n\}$, corresponding to the conserved quantities $Q^{(n)}_L$ and $Q^{(n)}_R$. Substituting the expression for $Q^{(n)}$ from Eq.~\eqref{def:Q^n} into Eq.~\eqref{icLR}, 
we can express $P_{\rm in}$ in the following Gaussian form
\begin{align}
P_{\rm in} =\frac{1}{Z^L Z^R} \exp\left(-\displaystyle{\frac{1}{2}} [\mathbf{p}^{\text{T}} \mathcal{B} \mathbf{p} + \mathbf{q}^{\text{T}} \mathcal{M} \mathbf{q}]\right), \label{ic_domainWall}
\end{align}
where $\mathcal{B}$ and $\mathcal{M}$ are $2N \times 2N$ symmetric matrices defined as
\begin{align}
\mathcal{B} = 
\begin{pmatrix}
\sum\limits_n \lambda_n^L B^{(n)} & 0 \\
0 & \sum\limits_n \lambda_n^R B^{(n)}
\end{pmatrix}, ~~~~
\mathcal{M} = 
\begin{pmatrix}
\sum\limits_n \lambda_n^L M^{(n)} & 0 \\
0 & \sum\limits_n \lambda_n^R M^{(n)}
\end{pmatrix},  \label{BMmat}
\end{align}
where the Lagrange multipliers are chosen such that $\mathcal{B}$ and $\mathcal{M}$ are positive definite matrices. 

The initial correlations can be immediately inferred from Eq.~\eqref{ic_domainWall} and are given by  $\expval{\mathbf{q}(0) \mathbf{q}(0)^T}=\mathcal{M}^{-1}$, $\expval{\mathbf{p}(0) \mathbf{p}(0)^T}=\mathcal{B}^{-1}$, and $\expval{\mathbf{p}(0) \mathbf{q}(0)^T}=0$. Using Eq.~\eqref{q_t-p_t}, the equal time two-point correlations can then be easily obtained and we find 
\begin{subequations}
\begin{align}
\expval{\mathbf{q}(t) \mathbf{q}(t)^T} &= A(t) \mathcal{B}^{-1} A(t) + \dot{A}(t) \mathcal{M}^{-1} \dot{A}(t), \\
\expval{\mathbf{p}(t) \mathbf{p}(t)^T} &= \dot{A}(t) \mathcal{B}^{-1} \dot{A}(t) + \ddot{A}(t) \mathcal{M}^{-1} \ddot{A}(t),\\
\expval{\mathbf{q}(t) \mathbf{p}(t)^T} &= A(t) \mathcal{B}^{-1} \dot{A}(t) + \dot{A}(t) \mathcal{M}^{-1} \ddot{A}(t),
\end{align} 
\label{qq_pp_mat}
\end{subequations}
where $A(t)$ is given in Eq.~\eqref{A(t)}.

\section{Wigner function, conserved quantities and connection to GHD} \label{sec:wignerCQGHD}
In this section, we discuss the Generalized Hydrodynamics (GHD) for the harmonic chain through its connection to the Wigner function formalism. This is similar to the approach followed in \cite{essler2022} for non-interacting fermions on a one-dimensional lattice. Using GHD, we obtain an analytical understanding of the correlations and conserved quantities on a macroscopic scale. Unless stated otherwise, we will assume the chain to be infinite at both ends for which the eigen-spectrum in Eq.~\eqref{w_k} becomes
continuous: $\omega^2(k) = 2(1-\cos k)$ where $k\in [-\pi,\pi]$. 

\subsection{\textbf{Wigner function: its relation to the correlation matrix and evolution }} \label{subsec:wignerCorrMat}
Although the Wigner function was originally introduced as a quantum analogue of the classical phase space distribution \cite{wigner1932,hillery1997}, it has also been used widely in the context of transport and hydrodynamics of classical harmonic crystals. 
Following \cite{spohn2006boltzmann}, we construct the Wigner function for the classical harmonic chain defined in Eq.~\eqref{hamiltonian}. For this, we first define the quantity 
\begin{align}
a(k, \tau) = \displaystyle{\frac{1}{\sqrt{2}}} \left[ \sqrt{\omega(k)} q(k, \tau) + \frac{i}{\sqrt{\omega(k)}} p(k, \tau) \right], \label{adef}
\end{align}
where the Fourier and inverse Fourier transforms are defined as
\begin{subequations}
\begin{align}
q(k, \tau) &= \sum_{j \in \mathbb{Z}} e^{-ijk} q_j(\tau); ~~~ q_j(\tau) = \int_{-\pi}^{\pi} \displaystyle{\frac{dk}{2\pi}} e^{ijk} q(k, \tau), \\
p(k, \tau) &= \sum_{j \in \mathbb{Z}} e^{-ijk} p_j(\tau); ~~~p_j(\tau) = \int_{-\pi}^{\pi} \displaystyle{\frac{dk}{2\pi}} e^{ijk} p(k, \tau), \\
a(k, \tau) &= \sum_{j \in \mathbb{Z}} e^{-ijk} a_j(\tau); ~~~ a_j(\tau) = \int_{-\pi}^{\pi} \displaystyle{\frac{dk}{2\pi}} e^{ijk}a(k, \tau),
\end{align}
\label{fourierTransforms}
\end{subequations}
where $a_j(\tau)$ is defined as the Fouier transform of $a(k, \tau)$. The quantity $a(k, \tau)$ has a simple evolution: $a(k, \tau) = e^{-i \omega(k) \tau} a(k, 0)$. We then define the Wigner function in terms of $a(k, \tau)$ as
\begin{align}
W^1(y, k, \tau) &= \int_{-2 \pi}^{2 \pi} \displaystyle{\frac{d\xi}{4 \pi}}  e^{i \xi y} \expval{a^{\star}(k - \xi/2, \tau) a(k + \xi/2, \tau)} = \sum_{j \in \mathbb{Z}} e^{-i j k} \expval{a^{\star}_{y - j/2}(\tau) a_{y + j/2}(\tau)}, \label{wigner1}
\end{align}
where $y \in \mathbb{Z}/2$. The sum on $j$ runs over even/odd integers if $y$ is integer/half-integer and the average is over an initial ensemble. Substituting Eq.~\eqref{adef} into Eq.~\eqref{wigner1}, we obtain
\begin{align}
\begin{split}
W^1(y, k, \tau) &= \displaystyle{\frac{1}{2}} \int_{-2\pi}^{2\pi} \displaystyle{\frac{d \xi}{4 \pi}} \sum_{j, j' \in \mathbb{Z}} e^{i \xi (y - (j+j')/2)} e^{i(j - j')k}  \\ 
& \times~\left[ \sqrt{\omega(k - \xi/2) \omega(k + \xi/2)} \expval{q_j q_{j'}}(\tau) + i \sqrt{\frac{\omega(k - \xi/2)}{\omega(k + \xi/2)}} \expval{q_j p_{j'}}(\tau)  \right. \\
&~~\left. - i \sqrt{\frac{\omega(k + \xi/2)}{\omega(k - \xi/2)}} \expval{p_j q_{j'}}(\tau)  + \frac{1}{\sqrt{\omega(k - \xi/2) \omega(k + \xi/2)}} \expval{p_j p_{j'}}(\tau)  \right] .
\end{split} 
\label{wignerCorr}
\end{align}
Note that the above expression readily provides connection between the Wigner function and the correlations. 
If we assume the correlations to be homogeneous in space, then we obtain a steady-state Wigner function
\begin{align}
W_{ss}(k) = \displaystyle{\frac{1}{2}} \left[ \omega(k) F_{11}(k) - i F_{12}(k) + i F_{21}(k) + \displaystyle{\frac{1}{\omega(k)}} F_{22}(k) \right], \label{wignerDobrushin1}
\end{align}
where $F_{m n}(k)$ are elements of the $2 \times 2$ correlation matrix
\begin{align}
F(k) = \sum_{r\in \mathbb{Z}} e^{ikr}\bar{F}(r),~~\text{where}~~\bar{F}({j-j'}) = 
\begin{pmatrix}
\expval{q_j q_{j'}} & \expval{q_j p_{j'}} \\[5pt]
\expval{p_j q_{j'}} & \expval{p_j p_{j'}}
\end{pmatrix}.
\label{F_j'j}
\end{align} 
We point out that $F_{11}^{*}(k) = F_{11}(k),~ F_{22}^{*}(k) = F_{22}(k)$ and $F_{12}^{*}(k) = F_{21}(k)$ so that the Wigner function is always real. In thermal equilibrium at an inverse temperature $\beta$, we simply get
\begin{align}
W_{\beta}(k) = \displaystyle{\frac{1}{\beta \omega(k)}}. \label{wignerThermal}
\end{align}
In general, for a chain of $N$ particles in GGE described by the distribution in Eq.~\eqref{P_GGE}, the Wigner function can be obtained by computing the correlations $F_{j-j'}$ in Eq.~\eqref{F_j'j} and using them in Eq.~\eqref{wignerDobrushin1}. One finds the following explicit expression
\begin{align}
W_{GGE}(k) = \displaystyle{\frac{1}{\omega(k) \sum\limits_{n=0}^{N-1} \lambda_n \cos n k}}. \label{wignerGGE}
\end{align}

On the other hand, when the correlation functions are inhomogeneous and evolve over a large space-time scale, one can assume the system to be locally in a steady state characterized by the local values of the correlations. Transforming to the scaled variables: $x = \epsilon y, t = \epsilon \tau$, and taking  the $\epsilon \to 0$ limit  in Eq.~\eqref{wignerCorr}, the relation in Eq.~\eqref{wignerDobrushin1} between the Wigner function and the correlation matrix gets modified to 
\begin{align}
W(x,k,t) = \displaystyle{\frac{1}{2}} \left[ \omega(k) F_{11}(x,k,t) - i F_{12}(x,k,t) + i F_{21}(x,k,t) + \displaystyle{\frac{1}{\omega(k)}} F_{22}(x,k,t) \right], \label{wignerDobrushin2}
\end{align}
where 
\begin{align}
W(x, k, t)&:= \lim\limits_{\epsilon \to 0} W^1(x/\epsilon,k,t/\epsilon) = \lim\limits_{\epsilon \to 0} \epsilon \int_{-2 \pi/\epsilon}^{2 \pi/\epsilon} \displaystyle{\frac{d\eta}{4 \pi}}  e^{i \eta x} \expval{a^{\star}(k - \epsilon \eta/2, t/\epsilon) a(k + \epsilon \eta/2, t/\epsilon)}, \label{wigner2}
\end{align}
and
\begin{align}
F_{mn}(x,k,t) = \lim\limits_{\epsilon \to 0} \sum\limits_{r=-\infty}^{\infty} e^{-irk} \expval{y^{m}_{[\epsilon^{-1}x-r/2]} ~y^n_{[\epsilon^{-1}x+r/2]}}(\epsilon^{-1}t).
\label{wignerDobrushin3}
\end{align}
Here $y^1_j = q_j$ is the displacement and $y^2_j = p_j$ is the momentum of the $j^{\text{th}}$ oscillator. The local  correlation functions $F_{mn}(x,k,t)$ were shown, in Ref.~\cite{dobrushin1986}, to satisfy the Euler equations, using which one can show that the Wigner function $W(x,k,t)$ satisfies the following transport equation:
\begin{align}
\partial_t W(x, k, t) + \omega'(k) \partial_x W(x, k, t) = 0. \label{wignerEqn}
\end{align}
A more direct way to derive this equation is to start with  $W^1(y,k,\tau)$ defined in Eq.~\eqref{wigner1} which satisfies
\begin{align}
\partial_\tau W^1(y,k,\tau) = i \int_{-2 \pi}^{2 \pi} \displaystyle{\frac{d\xi}{4 \pi}}  e^{i \xi y}[\omega(k-\xi/2) - \omega(k+\xi/2)] \expval{a^{\star}(k - \xi/2, \tau) a(k + \xi/2, \tau)}. \label{W^1-evo}
\end{align}
We then transform to the scaled variables $(x=\epsilon y, t = \epsilon \tau)$ on both sides of Eq.~\eqref{W^1-evo} and expand the right-hand side in $\epsilon$. At the leading order in $\epsilon$, one finds the evolution equation \eqref{wignerEqn} for $W(x,k,t)$ defined in Eq.~\eqref{wigner2}.

On an infinite line, the solution of Eq.~\eqref{wignerEqn} at a later time can be simply obtained by boosting the initial function $W(x, k, 0)$ with the phonon group velocity $\omega'(k)$:
\begin{align}
W(x, k, t) = W(x-\omega'(k)t, k ,0). \label{wignerSoln}
\end{align}


\noindent
Note that in thermal equilibrium the average internal energy at a site can be obtained from $W_{\beta}(k)$ [given in Eq.~\eqref{wignerThermal}] as $\int_{-\pi}^{\pi} dk/(2 \pi) \omega(k) W_{eq}(k) = 1/\beta$. Generalizing this relation to the inhomogeneous case we find that the average internal energy density can be expressed in terms of the Wigner function as
\begin{align}
e(x, t) = \int_{-\pi}^{\pi} \displaystyle{\frac{dk}{2\pi}} \omega(k) W(x, k, t). \label{energyWigner}
\end{align}
It is easy to see that $e(x,t)$ satisfies a continuity equation $\partial_t e(x, t) + \partial_x \mathfrak{j}(x, t) = 0$ where the average energy current density is given by
\begin{align}
\mathfrak{j}(x, t) = \int_{-\pi}^{\pi} \displaystyle{\frac{dk}{2\pi}} \omega(k) \omega'(k) W(x, k, t). \label{currentdef}
\end{align} 
In the following, we discuss other conserved densities and the associated currents and show how one can express them in terms of the Wigner function.

\subsection{\textbf{Conserved quantities and connection to GHD}} \label{subsec:conservedGHD}
The Hamiltonian in Eq.~\eqref{hamiltonian} for $N \to \infty$ can be written in Fourier space as 
\begin{align}
H = \bigintsss\limits_{-\pi}^{\pi} \displaystyle{\frac{dk}{2 \pi}} \mathcal{E}(k)~~\text{with}~~\mathcal{E}(k)=  \displaystyle{\frac{\omega(k)}{2}} [a^{\star}(k, \tau) a(k, \tau) + a^{\star}(-k, \tau) a(-k, \tau)],
\end{align}
where $a(k,\tau)$ is defined in Eq.~\eqref{adef}. Note that $\mathcal{E}(k)= {\frac{|p(k, \tau)|^2}{2}} + \omega(k)^2 \frac{|q(k, \tau)|^2}{2}$ represents the energy of a single mode and is a conserved quantity. 
This allows us to define a set of new conserved quantities as the following linear combinations:
\begin{align}
Q^{(n)} := \int\limits_{-\pi}^{\pi} \displaystyle{\frac{dk}{2 \pi}} \mathcal{E}(k) e^{ink}= \int\limits_{-\pi}^{\pi} \displaystyle{\frac{dk}{2 \pi}} \cos nk ~\omega(k) a^{\star}(k, \tau) a(k, \tau) ~~\text{for}~n=0,1,2,... \label{Qkspace2}
\end{align}
We now show that these conserved quantities are local in nature, i.e., they can be expressed as a sum over local terms. 
To see this,  we  insert Eq.~\eqref{adef} into Eq.~\eqref{Qkspace2} and after  some simplifications, obtain
\begin{align}
Q^{(n)} = \sum_{\ell \in \mathbb{Z}} \varrho^{(n)}_{\ell}(\tau)~~\text{where}~~
\varrho^{(n)}_{\ell}(\tau)=\displaystyle{\frac{1}{2}}\left[ p_{\ell} p _{\ell-n} + 2q_{\ell} q_{\ell-n} - q_{\ell} q_{\ell-n+1} - q_{\ell} q_{\ell-n-1} \right]. \label{local-Q_n}
\end{align}
It is easy to see that the microscopic densities $\varrho^{(n)}_{\ell}(\tau)$ obey a discrete continuity equation
\begin{align}
\partial_{\tau}\varrho^{(n)}_{\ell}(\tau) = \mathfrak{j}^{(n)}_{\ell-1}(\tau) - \mathfrak{j}^{(n)}_{\ell}(\tau), \label{continuityEqn2}
\end{align}
where the microscopic current $\mathfrak{j}^{(n)}_{\ell}(\tau)$ from site $\ell$ to $\ell+1$ is given by 
\begin{align}
\mathfrak{j}^{(n)}_{\ell}(\tau) = \displaystyle{\frac{1}{2}} \left( p_{\ell-n+1} q_{\ell} - p_{\ell-n} q_{\ell+1} \right). \label{currMicro}
\end{align}
Note that  $Q^{(n)}$ in Eq.~\eqref{local-Q_n} can also  be written in matrix form as
\begin{align}
Q^{(n)} = \displaystyle{\frac{1}{2}} \left( \mathbf{p}^T B^{(n)} \mathbf{p} + \mathbf{q}^T M^{(n)} \mathbf{q} \right), \label{Qnmat}
\end{align}
where $B^{(n)}$ and $M^{(n)}$ are symmetric matrices defined as
\begin{subequations}
\begin{align}
B^{(n)}_{\ell, \ell'} &= \frac{1}{2} \left( \delta_{\ell-n, \ell'} + \delta_{\ell+n, \ell'} \right), \label{Bndef}\\
M^{(n)}_{\ell, \ell'} &= \delta_{\ell-n, \ell'} + \delta_{\ell+n, \ell'} - \frac{1}{2}\left( \delta_{\ell-n+1, \ell'} + \delta_{\ell+n-1, \ell'} + \delta_{\ell-n-1, \ell'} + \delta_{\ell+n+1, \ell'} \right). \label{Mndef}
\end{align} 
\label{Bn_Mn_mat}
\end{subequations}

\noindent These matrix forms are particularly useful for numerical microscopic computations. For initial conditions chosen from a distribution such as in Eq.~\eqref{icLR}, one is interested in the average values of these conserved densities $\rho_{\ell}^{(n)} := \expval{\varrho_{\ell}^{(n)}}$, which, on a macroscopic scale can be expressed in terms of the coarse-grained Wigner function $W(x,k,t)$. Noting $\expval{a^*(k,\tau)a(k,\tau)} = \sum_{y}W^1(y,k,\tau)$ 
from Eq.~\eqref{wigner1} and substituting it in Eq.~\eqref{Qkspace2} gives
\begin{align}
\expval{Q^{(n)}}=\sum_y\int\limits_{-\pi}^{\pi} \displaystyle{\frac{dk}{2 \pi}} \cos nk ~\omega(k) W^1(y,k,\tau).
\end{align}
Defining $\rho^{(n)}(x,t) = \lim_{\epsilon \to 0} \epsilon^{-1} \rho^{(n)}_{[\epsilon^{-1}x]}(\epsilon^{-1}t)$, we can write $\expval{Q^{(n)}}=\int dx \rho^{(n)}(x, t)$, where 
\begin{align}
\rho^{(n)}(x, t) = \int\limits_{-\pi}^{\pi} \displaystyle{\frac{dk}{2 \pi}} \cos nk ~\omega(k) W(x, k, t). \label{nrho}
\end{align}


\noindent Using Eq.~\eqref{wignerEqn}, it is easy to see that the densities in Eq.~\eqref{nrho} obey the continuity equation
\begin{align}
\partial_t \rho^{(n)}(x, t) + \partial_x \mathfrak{j}^{(n)}(x, t) = 0, \label{continuityEqn1}
\end{align} 
where the macroscopic currents $\mathfrak{j}^{(n)}(x, t)$ are given by 
\begin{align}
\mathfrak{j}^{(n)}(x, t) = \int\limits_{-\pi}^{\pi} \displaystyle{\frac{dk}{2 \pi}} \cos nk ~\omega(k) \omega'(k) W(x, k, t). \label{ncurrent}
\end{align}

We now comment on the connection to GHD. We see that the coarse-grained Wigner function in Eq.~\eqref{wigner2} indeed has the structure of GHD equations if we identify $W(x,k,t)$ as a phase space distribution of non-interacting quasi-particles (basically the phonons) with position $x$, momentum $k$, and velocity $\omega'(k)$. This identification is natural as was pointed out in \cite{spohn2006boltzmann}. The definition of conserved densities and currents in Eqs.~(\ref{nrho}, \ref{ncurrent}) is consistent with this identification. One question is regarding the positivity of $W$. At the microscopic level, it can be negative but coarse-graining can lead to a positive-definite quantity. Various possibilities for coarse-graining exist, for example through the construction of the Husimi function (see \cite{mielke2006}). Another more physical construction was discussed in \cite{pandey2023} in the context of a non-interacting quantum gas where the coarse-grained Wigner function has the direct interpretation as the number of quasi-particles in a coarse-grained cell. Note that at a physical level, the GHD equation for the harmonic chain is the same as the Peierls-Boltzmann equation for a weakly anharmonic chain with collision terms neglected (see \cite{simoncelli2022}).

\section{Comparison between microscopic and hydrodynamic evolution from domain wall initial condition} \label{sec:domainWall}
In this section, we apply the hydrodynamic formulation developed in Sec.~\eqref{sec:wignerCQGHD} to the study of equilibration of the harmonic chain starting from a domain wall initial condition as depicted in Fig.~\eqref{schematic}. We study both a chain of infinite extent and a chain of finite length with fixed boundaries. For the hydrodynamics of a finite-sized chain, we assume its length to be $2L$ with $2N$ particles and lattice constant $\epsilon$ such that $N\to \infty$ and $\epsilon \to 0$ with $L = N \epsilon$ held fixed. The microscopic numerical calculations are done for finite $N$ with unit lattice constant and to compare the results, we make the identification $L=N$. For both cases, we consider two types of domain wall initial conditions. In the first choice, only the (inverse) temperatures in the two halves of the chain are different, namely $\beta_1$ and $\beta_2$, while all other Lagrange multipliers are zero {\it{i.e.,}} $\lambda^L_n=\lambda^R_n=0,~\forall~n\ge 1$. For a finite chain, this implies that the only conserved quantity with a non-zero average value initially is the energy. As a consequence of conservation laws, the values of higher conserved quantities remain zero in the final stationary state which is thus again expected to be described by a GE with the final temperature being the mean of the initial temperatures in the two halves.  In Sec.~\eqref{subsec:domainWallGE}, by studying the evolution of the conserved densities we show how the system relaxes to the final expected GE state.  
In Sec.~\eqref{subsec:domainWallGGE}, we consider the second choice of the initial condition in which the first two parameters (Lagrange multipliers) are non-zero and take different values in the two halves while other parameters are zero on both sides. In this case, all the conserved quantities have non-zero average values initially, and consequently, we expect the system to finally reach a GGE stationary state. Once again, by solving the GHD equation in Eq.~\eqref{wignerEqn}, we study the evolution of the conserved densities and the associated currents and show how the system in this case approaches the final expected GGE state. Note that for both types of initial conditions, the infinite chain is expected to go to a GGE stationary state which is current carrying. This is due to the presence of additional conserved quantities (see \cite{boldrighini1983conservedQ}).  

\subsection{Domain wall initial condition composed of two GE states} \label{subsec:domainWallGE}
As described above, initially, there is a domain wall at the center of the chain such that the left and the right halves are described by two GE states with unequal inverse temperatures $\beta_1$ and $\beta_2$ respectively. We first consider the simpler case of an infinite chain which will be followed by a discussion on finite-size chain. \\

\noindent \textbf{Infinite chain:} 
We assume the domain wall to be initially located at $x=0$. 
Using Eq.~\eqref{wignerThermal}, we can write down the Wigner function at $t=0$ as
\begin{align}
W(x, k, 0) = \displaystyle{\frac{1}{\omega(k)}} \left[ \frac{1}{\beta_1} - \left( \frac{1}{\beta_1} - \frac{1}{\beta_2} \right) \Theta \left( x \right) \right]. \label{wignerInfinite1} 
\end{align}
From Eq.~\eqref{wignerSoln} the  Wigner function at time $t$  can straightforwardly be written  as  
\begin{align}
W(x, k, t) = \displaystyle{\frac{1}{\omega(k)}} \left[ \frac{1}{\beta_1} - \left( \frac{1}{\beta_1} - \frac{1}{\beta_2} \right) \Theta \left( x - \omega'(k)t \right) \right]. \label{wignerInfinite2} 
\end{align}
Substituting Eq.~\eqref{wignerInfinite2} into Eq.~\eqref{energyWigner} and performing the integral, one can compute the average internal energy density $e(x,t)$. Since we assume the lattice spacing to be unity, the particle density is unity everywhere. Hence, $e(x,t)$ is essentially the temperature profile which is explicitly given by
\begin{align}
T(x, t) := e(x, t) = \begin{cases}
\displaystyle{\frac{1}{\beta_1}}, ~ x < -t \\ \\
\displaystyle{\frac{1}{2}} \left( \displaystyle{\frac{1}{\beta_1}} + \displaystyle{\frac{1}{\beta_2}} \right) - \displaystyle{\frac{1}{\pi}} \left( \displaystyle{\frac{1}{\beta_1}} - \displaystyle{\frac{1}{\beta_2}} \right) \sin^{-1} \left( \frac{x}{t} \right), ~ |x| < t \\ \\
\displaystyle{\frac{1}{\beta_2}}, ~ x > t
\end{cases} \label{tempInfinite2}
\end{align}
The expressions for other conserved quantities $\rho^{(n)}(x, t)$ for $n > 0$ can be calculated similarly and are given by 
\begin{align}
\rho^{(n)}(x, t) = \displaystyle{\frac{1}{2n \pi}} \left( \displaystyle{\frac{1}{\beta_1}} - \displaystyle{\frac{1}{\beta_2}} \right) \sin \left\{ 2 n \cos^{-1}\left(\frac{x}{t}\right) \right\} \Theta\left( 1-\frac{|x|}{t} \right). \label{nrhoInfinite}
\end{align}
Using the continuity equation in Eq.~\eqref{continuityEqn1} for $n=0$, we can also write down the energy current density
\begin{align}
\mathfrak{j}^{(0)}(x, t) = \displaystyle{\frac{1}{\pi}} \left( \displaystyle{\frac{1}{\beta_1}} - \displaystyle{\frac{1}{\beta_2}} \right) \sqrt{1 - \left(\frac{x}{t}\right)^2} \Theta\left( 1-\frac{|x|}{t}\right). \label{tempcurrInfinite}
\end{align}
The higher currents $\mathfrak{j}^{(n)}(x, t)$ for $n > 0$ can be similarly evaluated and one finds the following expression
\begin{align}
\mathfrak{j}^{(n)}(x, t) = \displaystyle{\frac{(-1)^{n+1}}{\pi (4 n^2-1)}} \left( \displaystyle{\frac{1}{\beta_1}} - \displaystyle{\frac{1}{\beta_2}} \right) \left[ 2 n \displaystyle{\frac{x}{t}} \sin \left(2n \mu\left(\frac{x}{t}\right)\right) + \sqrt{1 - \left( \frac{x}{t} \right)^2} \cos \left(2n \mu \left(\frac{x}{t}\right)\right) \right] \Theta\left( 1-\frac{|x|}{t} \right), 
\label{ncurrInfinite}
\end{align}
where $\mu(z) = \sin^{-1}\left(z\right)$. 

\begin{figure}[H]
\centering
\subfigure[]{
	\includegraphics[scale=0.47]{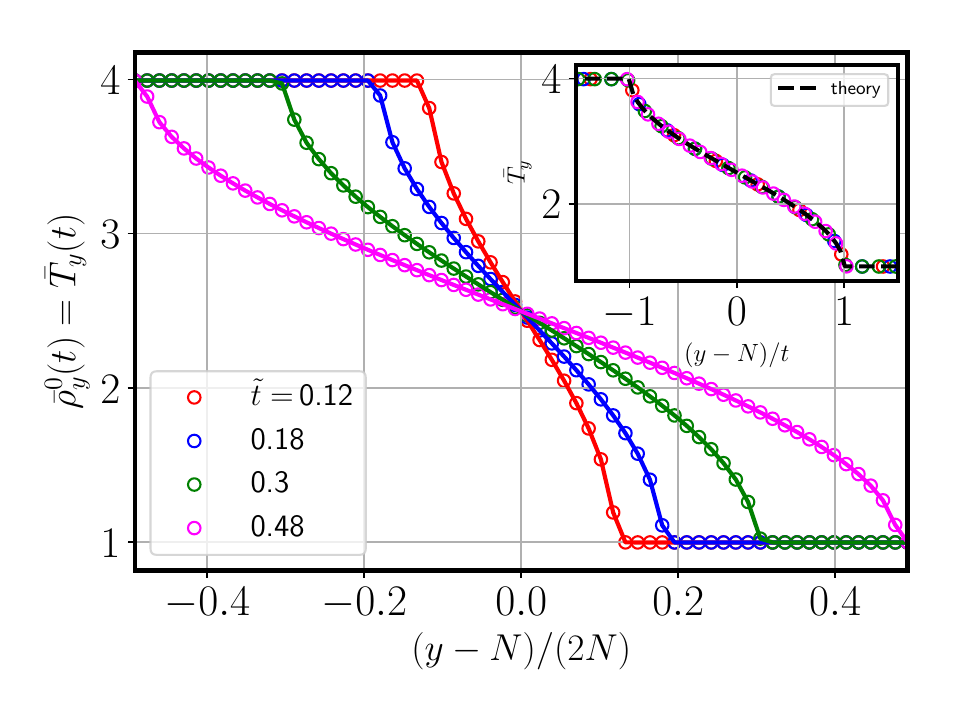}
}
\subfigure[]{
	\includegraphics[scale=0.47]{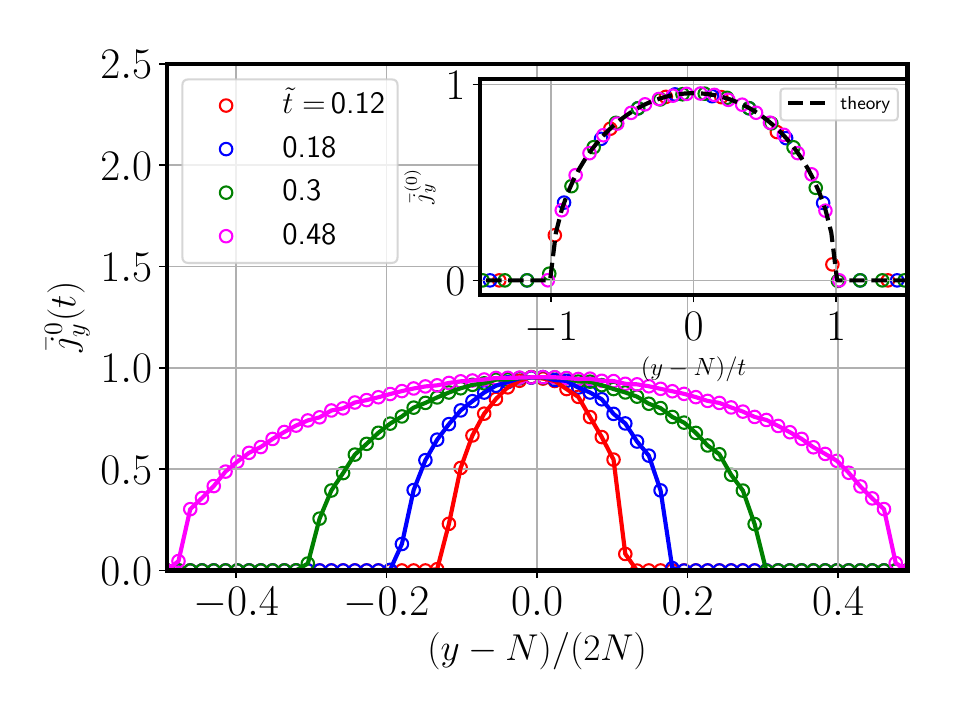}
}
\caption{Plot showing the early time behavior of the coarse-grained microscopic (a) temperature profile and (b) energy current density profile at different times (in the scaled units $\tilde{t}=t/(2N)$).  The insets in both figures show the collapse of the profiles upon $x/t$ scaling of the x-axis where $x = y-N$. The dashed black lines in the insets correspond to the analytic scaling functions obtained from  Eq.~\eqref{tempInfinite2} and Eq.~\eqref{tempcurrInfinite} respectively. The coarse-graining was done over $\Delta = 8$ consecutive sites. The parameters used are $\beta_1 = 0.25, \beta_2 = 1.0$ and  $2N=512$.} \label{1casecgQ0scaling}
\end{figure}

To compare the numerical microscopic computations with the theoretical results from hydrodynamics, we coarse-grain the microscopic densities (and currents) by averaging over $\Delta$ number of consecutive sites: 
\begin{align}
\bar{\rho}^{(n)}_y(t) = \displaystyle{\frac{1}{\Delta}} \sum\limits_{j = (\ell-1)\Delta + 1}^{\ell \Delta} \rho^{(n)}_j(t) ~~\text{where} ~~y=\ell \displaystyle{\frac{\Delta}{2}}.
\label{coarse-graining}
\end{align}
Note that we have expressed the coarse-grained densities as functions of macroscopic time variable $t$, because, as mentioned earlier, in our case $\epsilon=1$ which is small compared to $L=N$.
In Fig.~(\ref{1casecgQ0scaling}a) and Fig.~(\ref{1casecgQ0scaling}b), we plot the coarse-grained temperature and the corresponding current density obtained numerically from the microscopic computations in a chain of length $2N$. In the insets of these figures, we show that these profiles have ballistic scaling and we compare the scaling functions with those (dashed black lines) obtained analytically in Eqs.~(\ref{tempInfinite2}) and (\ref{tempcurrInfinite}) for temperature and the associated current respectively. 
The microscopic calculations are performed by evaluating averages of the corresponding microscopic quantities given in Eq.~\eqref{local-Q_n} and Eq.~\eqref{currMicro} and for that, we use the correlations from Eq.~\eqref{qq_pp_mat}. We see that the hydrodynamic results for both temperature and energy current density match with the corresponding finite-size microscopic computations as long as $t<N$. This can be understood since $N$ is the time taken by the step initial condition to spread and reach the boundaries. 

It is interesting to note that in the $t \to \infty$ limit, the currents at any fixed point $x$ in the system approach stationary nonzero values given by
\begin{align}
\mathfrak{j}^{(n)}_{ss} = \displaystyle{\frac{(-1)^{n+1}}{\pi (4 n^2-1)}} \left( \displaystyle{\frac{1}{\beta_1}} - \displaystyle{\frac{1}{\beta_2}} \right), \label{ncurrInfiniteGGE}
\end{align}
and the relaxation happens as $\sim 1/t^2$. On the other hand, at any fixed location in the $t \to \infty$ limit, all the densities of the conserved quantities approach zero except temperature which approaches the mean temperature. The relaxation to these stationary values occurs as $\sim 1/t$. Therefore any finite segment of the infinite chain reaches a non-equilibrium steady state (NESS) which cannot be described by a GE but possibly by a current carrying GGE instead, as shown for the quantum harmonic chain in \cite{eisler2014}.

As we will see below, this is different from the case of the finite chain where the steady state is indeed expected to be described by a GE since the currents, decay to zero due to reflections at the boundaries. 

\noindent \textbf{Finite chain:} We now turn our attention to the case of a finite chain. Let the length of the chain be $2L$ and the initial condition be that of a domain wall at $x=L$ as given in Eq.~\eqref{icLR} with $\lambda_0^{L} = \beta_1, \lambda_0^{R} = \beta_2$ and all other $\lambda_n^{L, R} = 0$. Using Eq.~\eqref{wignerThermal}, we can write down the Wigner function for the finite chain at $t=0$ as
\begin{align}
W(x, k, 0) = \displaystyle{\frac{1}{\omega(k)}} \left[ \frac{1}{\beta_2} + \left( \frac{1}{\beta_1} - \frac{1}{\beta_2} \right) \left[ \Theta \left( x + L \right) - \Theta \left( x - L \right) \right] \right], \label{wignerFiniteGE1}
\end{align}
where $x \in [0, 2L]$. Using Eq.~\eqref{wignerSoln} and taking care of reflections at the boundaries (following a similar procedure as done in Appendix A of Ref.~\cite{chakraborti2022}), the Wigner function at a later time $t$ is given by
\begin{align}
W(x, k, t) = \displaystyle{\frac{1}{\omega(k)}} \left[ \frac{1}{\beta_2} + \left( \frac{1}{\beta_1} - \frac{1}{\beta_2} \right) \sum_{s=-\infty}^{\infty} \left[ \Theta \left( x - \omega'(k)t + 4 s L + L \right) - \Theta \left( x - \omega'(k)t + 4 s L - L \right) \right] \right]. \label{wignerFiniteGE2}
\end{align}
Substituting Eq.~\eqref{wignerFiniteGE2} into Eq.~\eqref{nrho} with $n=0$, we obtain the following series representation  for the temperature
\begin{align}
T(x, t) = \displaystyle{\frac{1}{\beta_2}} + \left( \frac{1}{\beta_1} - \frac{1}{\beta_2} \right) \sum\limits_{s=-\infty}^{\infty} \left[ I \left( \frac{x+4sL+L}{t} \right) -  I \left( \frac{x+4sL-L}{t} \right) \right], \label{tempFiniteGE1}
\end{align}
where  we have defined the integral
\begin{align}
I(z) := \int\limits_{-1}^{1} \displaystyle{\frac{du}{\pi}} \frac{\Theta(z-u)}{\sqrt{1-u^2}} =  \begin{cases}
0, ~ z < -1 \\ \\
\displaystyle{\frac{1}{2}} + \frac{1}{\pi} \sin^{-1} z, ~ |z| < 1 \\ \\
1, ~ z > 1
\end{cases} \label{Iz_integral}
\end{align}
Note that Eq.~\eqref{wignerFiniteGE2} can alternatively be written in a compact form
\begin{align}
W(x, k, t) = \displaystyle{\frac{1}{\omega(k)}} \left[ \frac{1}{\beta_1} - \left( \frac{1}{\beta_1} - \frac{1}{\beta_2} \right) \Theta \left( \left\{ \displaystyle{\frac{x - \omega'(k)t + L}{4L}} \right\} - \frac{1}{2} \right) \right], \label{wignerFiniteGE3}
\end{align}
where $\{ x \} = x - \lfloor x \rfloor$ is the fractional part function.
Defining $z = \left\{ \displaystyle{\frac{x - \omega'(k)t + L}{4L}} \right\}$,  we  write the Fourier series representation, $\Theta \left( z - \displaystyle{\frac{1}{2}} \right) = \displaystyle{\frac{1}{2}} - \sum\limits_{\ell \neq 0} \frac{(e^{\pi i \ell} - 1) e^{-2 \pi i \ell z}}{2 \pi i \ell}$. Putting this in Eq.~\eqref{wignerFiniteGE3} gives
\begin{align}
W(x, k, t) = \displaystyle{\frac{1}{2 \omega(k)} \left( \frac{1}{\beta_1} + \frac{1}{\beta_2} \right)} + \frac{1}{\omega(k)} \left( \frac{1}{\beta_1} - \frac{1}{\beta_2} \right) \sum\limits_{\ell \neq 0} \frac{\sin (\pi \ell/2)}{(\pi \ell)} \exp \left( -2 \pi i \ell \frac{x - \omega'(k) t}{4L} \right). \label{wignerFiniteGE4}
\end{align}
Now substituting Eq.~\eqref{wignerFiniteGE4} into Eq.~\eqref{nrho} after setting $n=0$ and integrating over $k$ yields an alternate series representation for the temperature profile
\begin{align}
T(x, t) = \rho^{(0)}(x, t) =  \displaystyle{\frac{1}{2}} \left( \frac{1}{\beta_1} + \frac{1}{\beta_2} \right) + \left( \frac{1}{\beta_1} - \frac{1}{\beta_2} \right) \sum\limits_{\ell=1}^{\infty} \frac{\sin (\pi \ell/2)}{(\pi \ell/2)} \cos \frac{\pi \ell x}{2L} J_0 \left( \frac{\pi \ell t}{2L} \right), \label{tempFiniteGE2} 
\end{align}
where $J_{\nu}(z)$ is the Bessel function.  
From Eq.~\eqref{wignerFiniteGE4}, we can also compute all other  densities $\rho^{(n)}(x, t)$ for $n > 0$ and we find the following evolution 
\begin{align}
\rho^{(n)}(x, t) = (-1)^n \left( \frac{1}{\beta_1} - \frac{1}{\beta_2} \right) \sum\limits_{\ell=1}^{\infty} \frac{\sin (\pi \ell/2)}{(\pi \ell/2)} \cos \frac{\pi \ell x}{2L} J_{2n} \left( \frac{\pi \ell t}{2L} \right). \label{nrhoFinite}
\end{align} 
Similarly putting Eq.~\eqref{wignerFiniteGE4} into Eq.~\eqref{ncurrent} and integrating over $k$ gives us the current densities
\begin{subequations}
\begin{align}
\mathfrak{j}^{(0)}(x, t) &= \left( \frac{1}{\beta_1} - \frac{1}{\beta_2} \right) \sum\limits_{\ell=1}^{\infty} \frac{\sin (\pi \ell/2)}{(\pi \ell/2)} \sin \frac{\pi \ell x}{2L} J_1 \left( \frac{\pi \ell t}{2L} \right), \label{tempcurrFinite} \\
\mathfrak{j}^{(n)}(x, t) &= \displaystyle{\frac{(-1)^n}{2}} \left( \frac{1}{\beta_1} - \frac{1}{\beta_2} \right) \sum\limits_{\ell=1}^{\infty}  \frac{\sin (\pi \ell/2)}{(\pi \ell/2)} \sin \frac{\pi \ell x}{2L} \left[ J_{2n+1} \left( \frac{\pi \ell t}{2L} \right) - J_{2n-1} \left( \frac{\pi \ell t}{2L} \right) \right], ~n>0. \label{ncurrFinite}
\end{align} \label{currFinite}
\end{subequations}

\noindent The Bessel function $J_{\nu}(z)$  decays as $1/\sqrt{z}$ for large $z$. Therefore, the temperature goes to a steady state value of $(1/2)(\beta_1^{-1} + \beta_2^{-1})$ while all other quantities in Eqs.~(\ref{nrhoFinite}, \ref{currFinite}) go to zero. All quantities relax as $1/\sqrt{t}$ in the large $t$ limit. 

In Figs.~(\ref{1casecgQ0}-\ref{1casecgQ1}), we show a comparison of the analytic results from hydrodynamics, for temperature, energy current density, and $Q^{(1)}$ density, with those obtained numerically from the microscopics by computing averages of the corresponding microscopic quantities in  Eqs.~(\ref{local-Q_n}, \ref{currMicro}) and using the correlations from Eq.~\eqref{qq_pp_mat}. 

In Fig.~(\ref{1casecgQ0}a) we show the temperature profiles at different scaled times $\tilde{t}=t/2N$. The initial front spreads freely as long as $t<N$. {At $t=N$ the front reaches the boundaries for the first time and gets reflected back into the bulk. These reflections continue to even out the temperature in the two halves as can be observed from the profile corresponding to the largest time $\tilde{t}=9.68$ shown in (\ref{1casecgQ0}a).} The temperature eventually becomes flat and equal to the mean of the initial temperatures. In Fig.~(\ref{1casecgQ0}b), we show how the temperature evolves with time at a coarse-grained point $y\approx (2N)\times 0.46$ that is slightly to the left of the center. The temperature remains unchanged until $\tilde{t} = (N-y)/(2N) \approx 0.04$ which is the time taken by the edge of the front to reach that point (not visible in the scale of the plot). {After that it starts decaying as $\sim 1/t$ [see inset of Fig.~(\ref{1casecgQ0}b)] similar to the infinite chain case and continues to do so until an edge of the reflected front [in Fig.~(\ref{1casecgQ0}a)] crosses the particular coarse-grained point for the first time. The temperature at any given point continues to show an overall decay interspersed with the presence of isolated kinks that appear corresponding to the passing of an edge of the reflected front. Eventually, the temperature at a fixed coarse-grained point approaches the final value as $\sim 1/\sqrt{t}$ (as also expected from our hydrodynamic calculations).}
In both figures, points represent numerical data, and the solid lines are analytical results from Eq.~\eqref{tempFiniteGE2} and we observe an excellent agreement between the two, thus demonstrating the validity of the GHD results for the case of finite-size chains.

In Fig.~(\ref{1casecgQ0curr}a), we plot the energy current density profiles at a few different times. The current profile is initially zero everywhere and starts to grow from the center. As long as $t<N$, the current evolves according to Eq.~\eqref{tempcurrInfinite} after which the reflections from the boundaries become relevant. Fig.~(\ref{1casecgQ0curr}b) shows the evolution of energy current density at a coarse-grained point that lies near the middle of the left half. The current stays zero until $\tilde{t} = (N-y)/(2N) \approx 0.27$ which is the time taken by the edge of the front to reach that point. Here also we observe kinks that correspond to the passing of an edge of the reflected front of the current profile in Fig.~(\ref{1casecgQ0curr}a). {The current eventually approaches zero as $\sim 1/\sqrt{t}$ [see inset of Fig.~(\ref{1casecgQ0curr}b)].} 
We again see good agreement between {the numerical microscopic computations and the analytical results from hydrodynamics.  

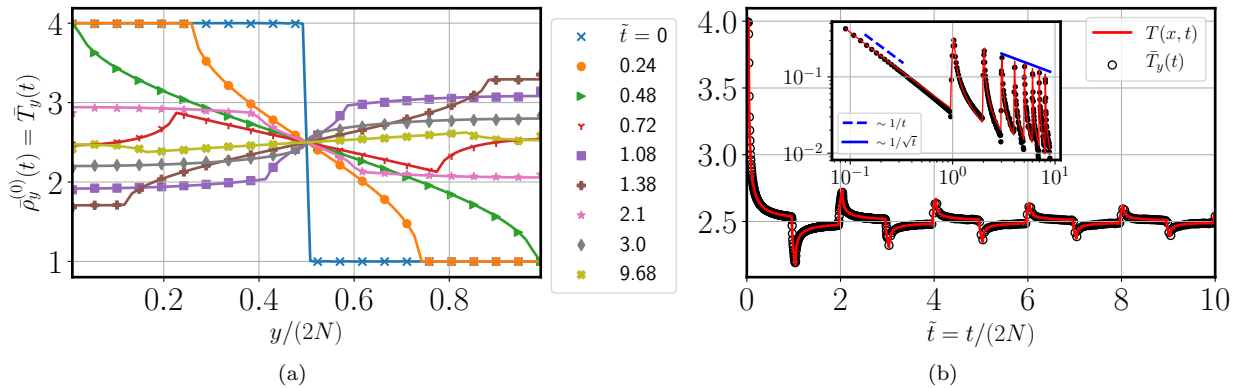
\begin{figure}[H]
\hspace*{-1.1cm}
\centering
    \includegraphics[scale=0.5]{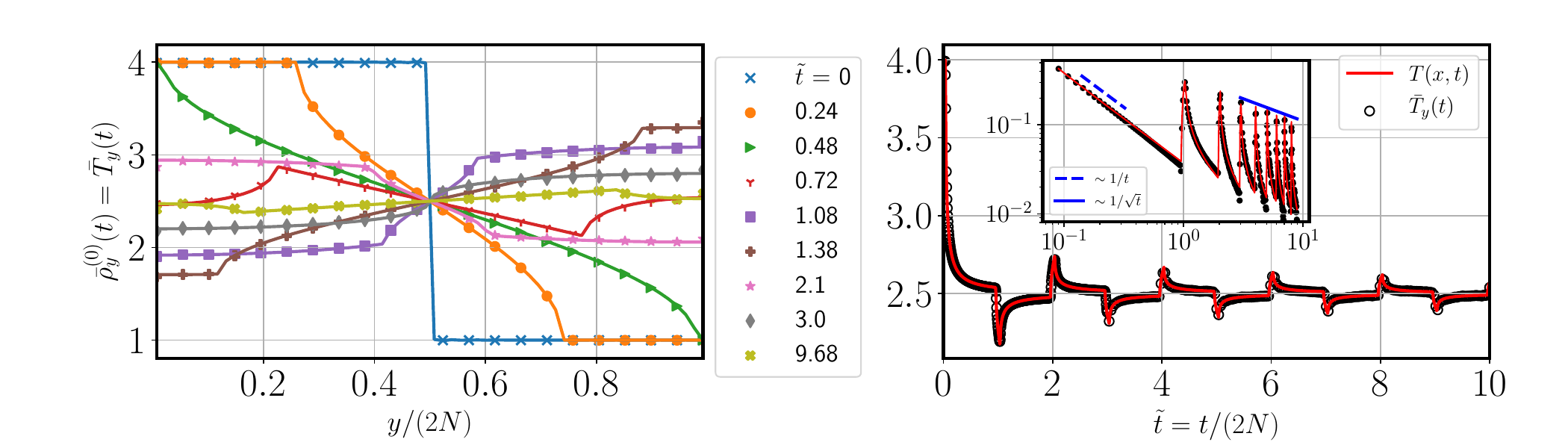}
\begin{tikzpicture}[>=latex]
    \node[anchor=east] at (2,0) {(a)}; 
    \node[anchor=west] at (10,0) {(b)};
\end{tikzpicture}
\caption{(a) Plot showing the coarse-grained temperature profile at different times  $\tilde{t}=t/(2N)$. (b) shows the evolution of the temperature with time at a fixed coarse-grained point $y/(2N) \approx 0.46$. The inset in (b) shows the same curve on a log-log scale but with the steady state value subtracted. We observe a $\sim 1/t$ decay at small times which, after many reflections from the boundaries, changes to a $\sim 1/\sqrt{t}$ decay at large times as expected from the analytical expressions. In both figures, the points are calculated by coarse-graining the microscopic temperature profile whereas the continuous lines are obtained from the analytic GHD solutions. We find an excellent agreement between the coarse-grained microscopic and the GHD results. The coarse-graining is done over $\Delta = 8$ consecutive sites. The parameters used are $\beta_1 = 0.25, \beta_2 = 1.0$ and $2N=512$.} \label{1casecgQ0}
\end{figure}

\begin{figure}[H]
\hspace*{-0.5cm}
\centering
    \includegraphics[scale=0.48]{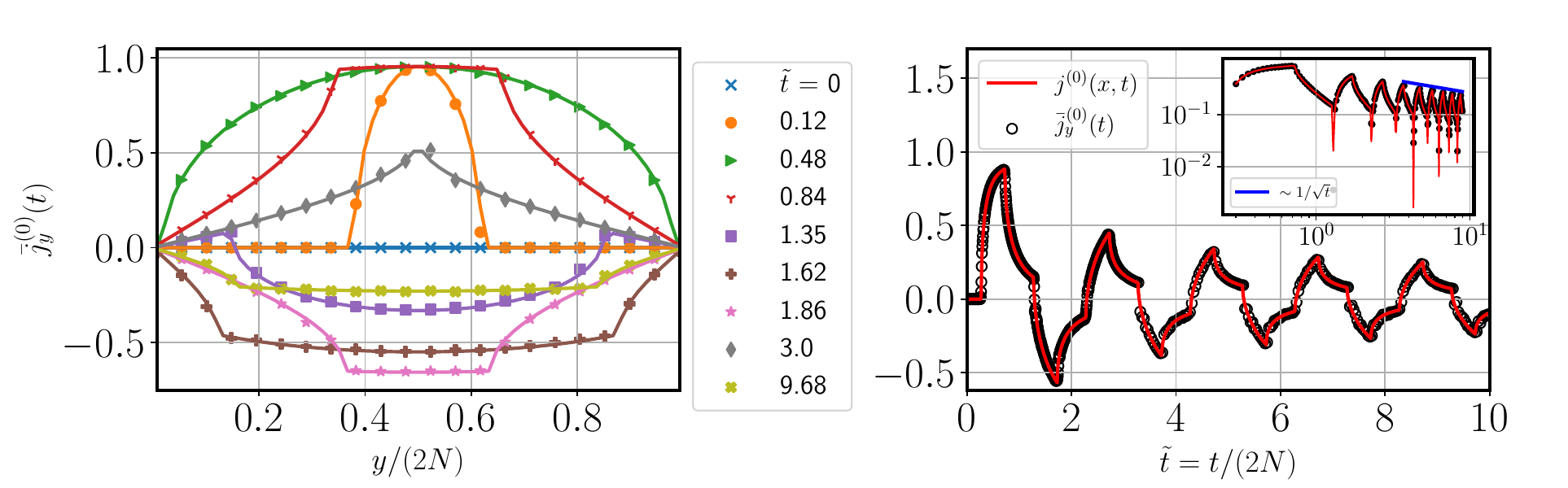}
\begin{tikzpicture}[>=latex]
    \node[anchor=east] at (3,0) {(a)}; 
    \node[anchor=west] at (11,0) {(b)};
\end{tikzpicture}
\caption{(a) Plot showing the coarse-grained energy current density profile at different times  $\tilde{t}=t/(2N)$. (b) shows the evolution of the energy current density with time at a fixed coarse-grained point $y/(2N) \approx 0.23$. The inset in (b) shows the same curve on a log-log scale. We observe a $\sim 1/\sqrt{t}$ decay at large times as expected from the analytical expressions. In both figures, the points are calculated by coarse-graining the microscopic energy current density profile whereas the continuous lines are obtained from the analytic GHD expressions. We find an excellent agreement between the coarse-grained microscopic and the GHD results. The coarse-graining is done over $\Delta = 8$ consecutive sites. The parameters used are $\beta_1 = 0.25, \beta_2 = 1.0$ and $2N=512$.} \label{1casecgQ0curr}
\end{figure}

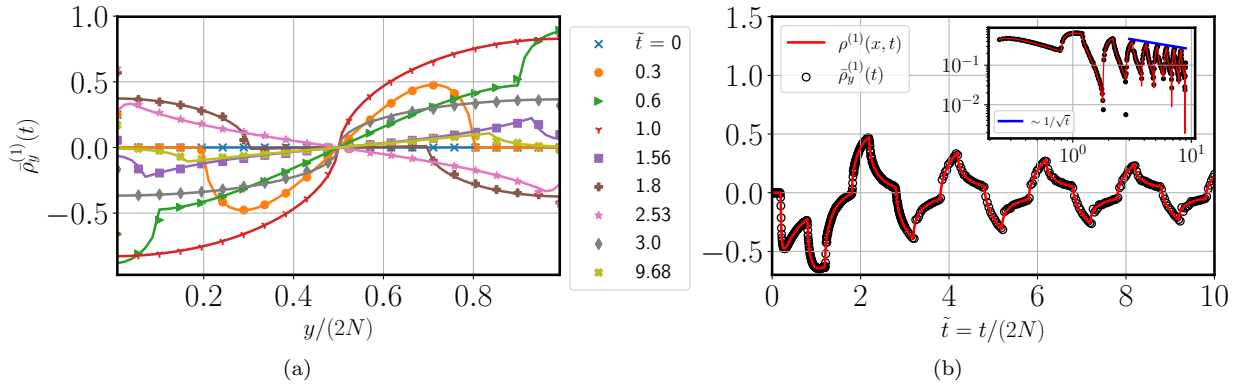
\begin{figure}[H]
\hspace*{-0.5cm}
\centering
    \includegraphics[scale=0.48]{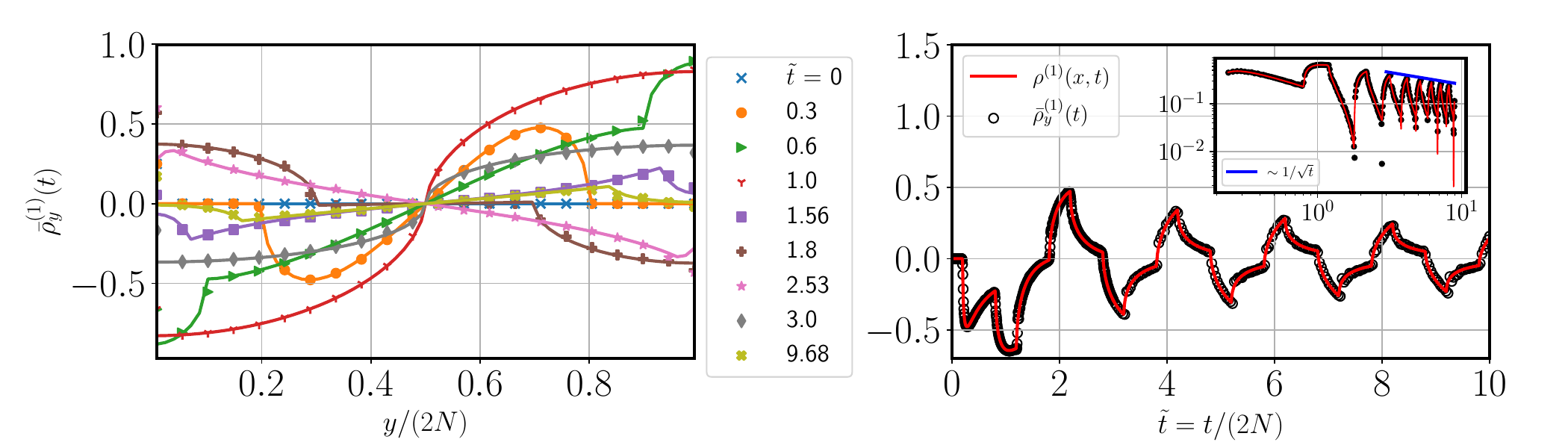}
\begin{tikzpicture}[>=latex]
    \node[anchor=east] at (2,0) {(a)}; 
    \node[anchor=west] at (10,0) {(b)};
\end{tikzpicture}
\caption{(a) Plot showing the coarse-grained $Q^{(1)}$ density profile at different times $\tilde{t}=t/(2N)$. (b) shows the evolution of the $Q^{(1)}$ density with time at a fixed coarse-grained point $y/(2N) \approx 0.30$. The inset in (b) shows the same curve on a log-log scale. We again observe a $\sim 1/\sqrt{t}$ decay at large times as expected from the analytical expressions. In both figures, the points are calculated by coarse-graining the microscopic $Q^{(1)}$ density profile whereas the continuous lines are obtained from the analytic GHD expressions. We find an excellent agreement between the coarse-grained microscopic and the GHD results. The coarse-graining is done over $\Delta = 8$ consecutive sites. The parameters used are $\beta_1 = 0.25, \beta_2 = 1.0$ and $2N=512$.} \label{1casecgQ1}
\end{figure}

\subsection{Domain wall initial condition composed of two GGE states} \label{subsec:domainWallGGE}
We now consider a more general initial condition where the two halves are described by a product of two GGEs, as given in Eq.~\eqref{icLR}. The initial Wigner functions in the left and right parts are given by Eq.~\eqref{wignerGGE} with the Lagrange parameters $\{ \lambda_n^L\}$ and $\{\lambda_n^R \}$ respectively. 
In the following, we focus on the particular case where only the first two parameters are non-zero and the rest are zero on both sides of the chain. \\

\noindent \textbf{Infinite chain:} Once again we assume that the domain wall is initially located at $x=0$ and the first two Lagrange multipliers have values 
$\lambda_0^L=\beta_1,~\lambda_1^L=-\beta_1\gamma_1,~\lambda_0^R=\beta_2$, and $\lambda_1^R=-\beta_2\gamma_2$. The initial Wigner function in this case is 
\begin{align}
W(x, k, 0) = \displaystyle{\frac{1}{\omega(k)}} \left[ \frac{1}{\beta_1(1 - \gamma_1 \cos k)} - \left( \frac{1}{\beta_1(1 - \gamma_1 \cos k)} - \frac{1}{\beta_2(1 - \gamma_2 \cos k)} \right) \Theta \left( x \right) \right], \label{wignerInfiniteGGE0}
\end{align}
which according to  Eq.~\eqref{wignerSoln} takes the following form at time $t$
\begin{align}
W(x, k, t) = \displaystyle{\frac{1}{\omega(k)}} \left[ \frac{1}{\beta_1(1 - \gamma_1 \cos k)} - \left( \frac{1}{\beta_1(1 - \gamma_1 \cos k)} - \frac{1}{\beta_2(1 - \gamma_2 \cos k)} \right) \Theta \left( x - \omega'(k)t \right) \right]. \label{wignerInfiniteGGE1}
\end{align} 
Using this solution in Eq.~\eqref{nrho} and simplifying we find that the density profiles associated with the conserved quantities are given by 
\begin{align}
\rho^{(n)}(x, t) = 
\begin{cases}
g^{(n)}(\beta_1, \gamma_1), ~ x < -t \\ \\
\displaystyle{\frac{1}{2}} [g^{(n)}(\beta_1, \gamma_1) + g^{(n)}(\beta_2, \gamma_2)] 
\\
- \bigintss\limits_{0}^{\sin^{-1}(x/t)} \displaystyle{\frac{d\theta}{\pi}} \cos \{ 2 n \cos^{-1} |\sin \theta| \} \left( \frac{1}{\beta_1(1 + \gamma_1 \cos 2\theta)} - \frac{1}{\beta_2(1 + \gamma_2 \cos 2\theta)} \right), ~ |x| < t \\ \\
g^{(n)}(\beta_2, \gamma_2), ~ x > t
\end{cases} \label{nrhoGGEInfinite2}
\end{align}
where the quantity $g^{(n)}(\beta, \gamma)$ is given by
\begin{align}
g^{(n)}(\beta, \gamma) = \displaystyle{\frac{1}{\beta (1 + \gamma)}} {}_3F^{\text{reg}}_{2} \left( \{1/2, 1, 1\}; \{1-n, 1+n\}; \displaystyle{\frac{2 \gamma}{1 + \gamma}} \right), \label{ngdef}  
\end{align}
with ${}_pF^{\text{reg}}_{q} (\{ a_1, \ldots, a_p \}; \{ b_1, \ldots, b_q \}; z)$ being the regularized generalized Hypergeometric function. The details of the calculation are given in Appendix~\eqref{app:GGE}. Putting $n=0$ in Eq.~\eqref{nrhoGGEInfinite2} and performing the integrals we find that the temperature profile is given by
\begin{align}
T(x, t) = \rho^{(0)}(x, t) =  
\begin{cases}
g^{(0)}(\beta_1, \gamma_1), & x < -t \\ \\
\displaystyle{\frac{1}{2}} \left( g^{(0)}(\beta_1, \gamma_1) + g^{(0)}(\beta_2, \gamma_2) \right) - \displaystyle{\frac{1}{\pi}} \left[ g^{(0)}(\beta_1, \gamma_1) \tan^{-1} \left\{ u(\gamma_1) s(x/t) \right\} \right. \\
\left. - g^{(0)}(\beta_2, \gamma_2) \tan^{-1} \left\{ u(\gamma_2) s(x/t) \right\} \right], & |x| < t \\ \\
g^{(0)}(\beta_2, \gamma_2), & x > t
\end{cases} \label{tempGGEInfinite}
\end{align}
where $g^{(0)}(\beta, \gamma) = \frac{1}{\beta \sqrt{1 - \gamma^2}}$ and $s(z)$ and $u(\gamma)$ are defined as
\begin{align}
s(z) = \frac{z}{\sqrt{1-z^2}}, ~ u(\gamma) = \sqrt{\displaystyle{\frac{1-\gamma}{1+\gamma}}}. \label{s_u_def}
\end{align}
The expression for the density corresponding to $Q^{(1)}$ is similarly given by
\begin{align}
\rho^{(1)}(x, t) = 
\begin{cases}
g^{(1)}(\beta_1, \gamma_1), & x < -t \\ \\
\displaystyle{\frac{1}{2}} \left( g^{(1)}(\beta_1, \gamma_1) + g^{(1)}(\beta_2, \gamma_2) \right) \\ 
- \displaystyle{\frac{1}{\pi}} \left[ g^{(1)}(\beta_1, \gamma_1) \tan^{-1} \left\{ u(\gamma_1) s(x/t) \right\} - g^{(1)}(\beta_2, \gamma_2) \tan^{-1} \left\{ u(\gamma_2) s(x/t) \right\} \right]  & |x| < t
\\
+ \displaystyle{\frac{1}{\pi}} \left[ \left( \displaystyle{\frac{1}{\beta_1 \gamma_1}} \tan^{-1} \left\{ \displaystyle{\frac{[1 - u(\gamma_1)]s(x/t)}{1+u(\gamma_1) s(x/t)^2}} \right\}  - \frac{1}{\beta_2 \gamma_2} \tan^{-1} \left\{ \displaystyle{\frac{[1 - u(\gamma_2)]s(x/t)}{1+u(\gamma_2) s(x/t)^2}} \right\} \right) \right],  \\ \\
g^{(1)}(\beta_2, \gamma_2),  & x > t
\end{cases} \label{1rhoGGEInfinite}
\end{align}
where $g^{(1)}(\beta, \gamma) = \frac{1}{\beta \gamma} \left( \frac{1}{\sqrt{1 - \gamma^2}} - 1 \right)$ and $s(z)$ and $u(\gamma)$ are defined in Eq.~\eqref{s_u_def}.

In Fig.~\eqref{2casecgQ0scaling}, we show a comparison of the analytic results from hydrodynamics with those obtained microscopically for a chain of length $2N$. The averages of the corresponding microscopic quantities in Eq.~\eqref{local-Q_n} are evaluated by using the correlations from Eq.~\eqref{qq_pp_mat}. We see that the hydrodynamic results in Eqs.~(\ref{tempGGEInfinite}) and (\ref{1rhoGGEInfinite}) for temperature and $Q^{(1)}$ density, respectively,  match with the corresponding finite-size microscopic computations as long as $t<N$.

\begin{figure}[H]
\centering
\subfigure[]{
	\includegraphics[scale=0.47]{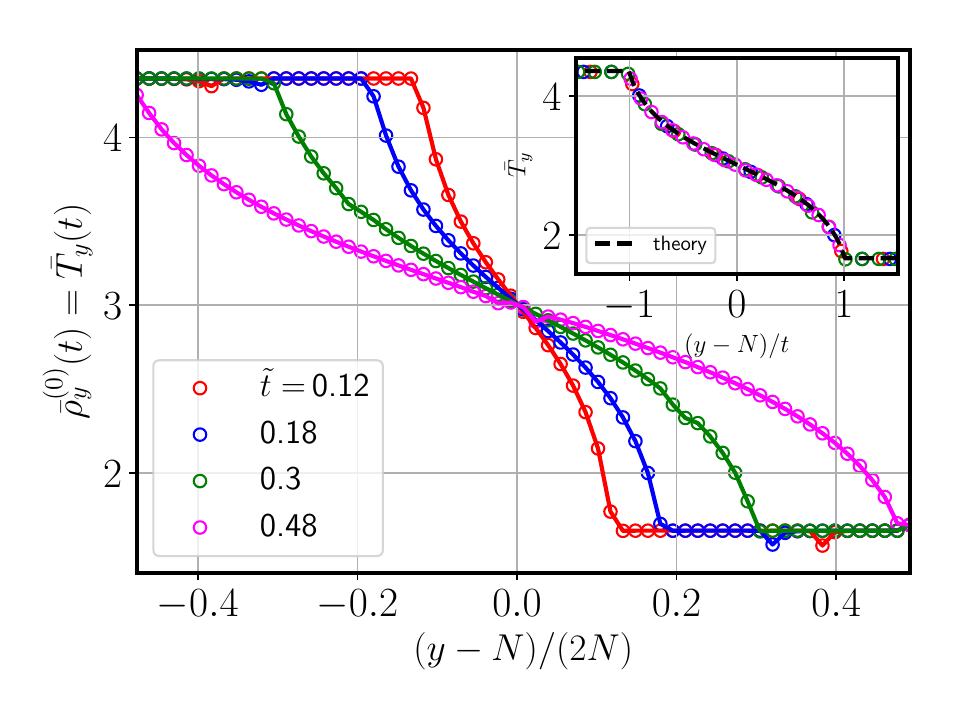}
}
\subfigure[]{
	\includegraphics[scale=0.47]{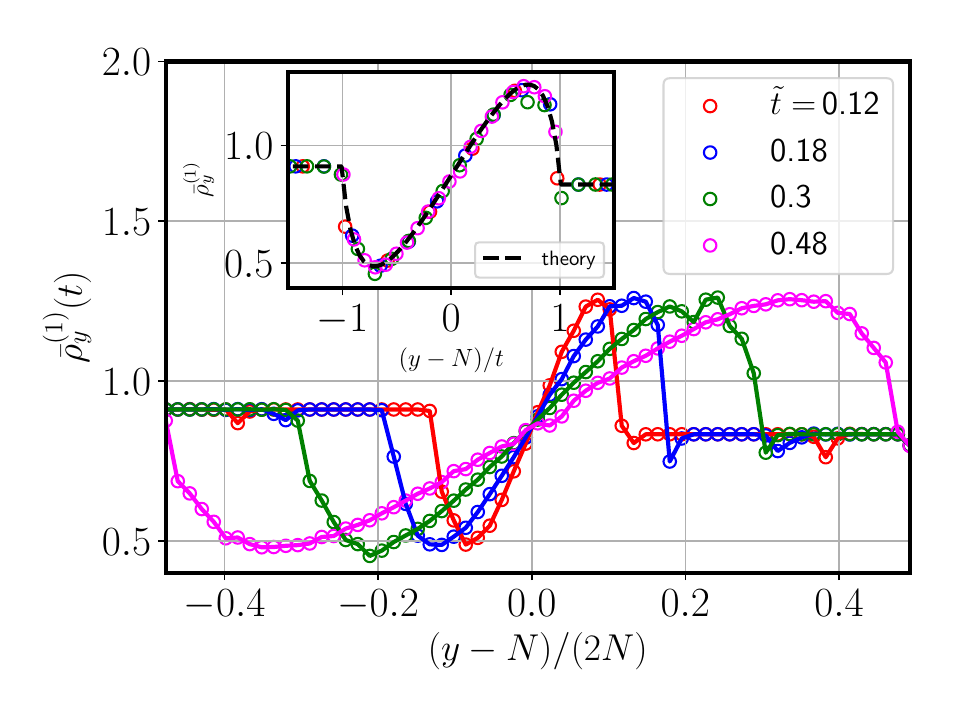}
}
\caption{Plot showing the early time behavior of the coarse-grained microscopic (a) temperature profile and (b) $Q^{(1)}$ density profile at different times (in the scaled units $\tilde{t}=t/(2N)$).  The insets in both figures show the collapse of the profiles upon $x/t$ scaling of the x-axis where $x = y-N$. The dashed black lines in the insets correspond to the analytic scaling functions obtained from  Eq.~\eqref{tempGGEInfinite} and Eq.~\eqref{1rhoGGEInfinite} respectively. The coarse-graining was done over $\Delta = 8$ consecutive sites. The parameters used are $\beta_1 = 0.25, \beta_2 = 1.0, \gamma_1 = 0.4, \gamma_2 = 0.8$ and  $2N=512$.} \label{2casecgQ0scaling}
\end{figure}

\noindent \textbf{Finite chain:} We now turn our attention to the case of a finite chain. Let the length of the chain be $2L$ and the initial condition be that of a domain wall at $x=L$ as given in Eq.~\eqref{icLR} with $\lambda_0^{L} = \beta_1, \lambda_1^{L} = -\beta_1 \gamma_1, \lambda_0^{R} = \beta_2, \lambda_1^{R} = -\beta_2 \gamma_2$ and all other $\lambda_n^{L, R} = 0$. Using Eq.~\eqref{wignerGGE} with these parameters in the two halves of the chain, we can write down the Wigner function at $t=0$ as
\begin{align}
W(x, k, 0) = \displaystyle{\frac{1}{\omega(k)}} \left[ \frac{1}{\beta_2(1 - \gamma_2 \cos k)} + \left( \frac{1}{\beta_1(1 - \gamma_1 \cos k)} -  \frac{1}{\beta_2(1 - \gamma_2 \cos k)} \right) \left[ \Theta \left( x + L \right) - \Theta \left( x - L \right) \right] \right], \label{wignerFiniteGGE1} 
\end{align}
where $x \in [0, 2L]$. As was argued in Sec.~\eqref{subsec:domainWallGE} for the finite chain, we can similarly write down the Wigner function at time $t$ as 
\begin{align}
\begin{split}
W(x, k, t) = \displaystyle{\frac{1}{\omega(k)}} &\left[ \frac{1}{\beta_2(1 - \gamma_2 \cos k)} + \left( \frac{1}{\beta_1(1 - \gamma_1 \cos k)} -  \frac{1}{\beta_2(1 - \gamma_2 \cos k)} \right) \right. \\
&\left. \times \sum_{s=-\infty}^{\infty} \left[ \Theta \left( x - \omega'(k)t + 4 s L + L \right) - \Theta \left( x - \omega'(k)t + 4 s L - L \right) \right] \right]. 
\end{split} \label{wignerFiniteGGE2} 
\end{align}
Substituting Eq.~\eqref{wignerFiniteGGE2} into Eq.~\eqref{nrho} with $n=0$, we obtain the following series representation for the temperature
\begin{align}
T(x, t) = \displaystyle{\frac{1}{\beta_2 \sqrt{1-\gamma_1^2}}} + \sum\limits_{s=-\infty}^{\infty} \left[ \frac{1}{\beta_1} I_{\gamma_1} \left( \frac{x+4sL+L}{t} \right) -  \frac{1}{\beta_2} I_{\gamma_2} \left( \frac{x+4sL-L}{t} \right) \right], \label{tempFiniteGGE1}
\end{align}
where $I_{\gamma}(z)$ is defined as the integral
\begin{align}
I_{\gamma}(z) := \int\limits_{-1}^{1} \displaystyle{\frac{du}{\pi}} \frac{\Theta(z-u)}{\sqrt{1-u^2} [1-\gamma (2u^2-1)]}. \label{I_gamma_z}
\end{align}
Note that, as before, Eq.~\eqref{wignerFiniteGGE2} can alternatively be written in a compact form
\begin{align}
W(x, k, t) = \displaystyle{\frac{1}{\omega(k)}} \left[ \frac{1}{\beta_1(1 - \gamma_1 \cos k)} - \left( \frac{1}{\beta_1(1 - \gamma_1 \cos k)} - \frac{1}{\beta_2(1 - \gamma_2 \cos k)} \right) \Theta \left( \left\{ \displaystyle{\frac{x - \omega'(k)t + L}{4L}} \right\} - \frac{1}{2} \right) \right], \label{wignerFiniteGGE3}
\end{align}
As done in Sec.~\eqref{subsec:domainWallGE} for the finite case, we define $z = \left\{ \displaystyle{\frac{x - \omega'(k)t + L}{4L}} \right\}$ and write the Fourier series representation, $\Theta \left( z - \displaystyle{\frac{1}{2}} \right) = \displaystyle{\frac{1}{2}} - \sum\limits_{\ell \neq 0} \frac{(e^{\pi i \ell} - 1) e^{-2 \pi i \ell z}}{2 \pi i \ell}$. Putting this in Eq.~\eqref{wignerFiniteGGE3} gives
\begin{align}
\begin{split}
W(x, k, t) &= \displaystyle{\frac{1}{2 \omega(k)} \left( \frac{1}{\beta_1(1 - \gamma_1 \cos k)} + \frac{1}{\beta_2(1 - \gamma_2 \cos k)} \right)} \\
&+ \frac{1}{\omega(k)} \left( \frac{1}{\beta_1(1 - \gamma_1 \cos k)} - \frac{1}{\beta_2(1 - \gamma_2 \cos k)} \right) \sum\limits_{\ell \neq 0} \frac{\sin (\pi \ell/2) \exp \left( -2 \pi i \ell \frac{x - \omega'(k) t}{4L} \right)}{\pi \ell}.
\end{split} \label{wignerFiniteGGE4}
\end{align}
Substituting Eq.~\eqref{wignerFiniteGGE4} into Eq.~\eqref{nrho}, the evolution for $\rho^{(n)}(x, t)$ can be formally written down as
\begin{align}
\begin{split}
\rho^{(n)}(x, t) &= \rho^{(n)}_{GGE} + \sum\limits_{\ell=1}^{\infty} \frac{\sin (\pi \ell/2)}{(\pi \ell/2)} \cos \frac{\pi \ell x}{2L} \left[ \displaystyle{\frac{1}{\beta_1}} C^{(n)}_{\ell}(\gamma_1, t) - \displaystyle{\frac{1}{\beta_2}} C^{(n)}_{\ell}(\gamma_2, t) \right],
\end{split} \label{nrhoGGE} 
\end{align}
where the stationary state density $\rho^{(n)}_{GGE}$ is given by
\begin{align}
\rho^{(n)}_{GGE} = \displaystyle{\frac{1}{2}} \left[ g^{(n)}(\beta_1, \gamma_1) + g^{(n)}(\beta_2, \gamma_2) \right], \label{GGEdensity}
\end{align}
with $g^{(n)}(\beta, \gamma)$  defined in Eq.~\eqref{ngdef} and $C^{(n)}_{\ell}(\gamma, t)$ given by the integral
\begin{align}
C^{(n)}_{\ell}(\gamma, t) = \int\limits_{0}^{\pi} \displaystyle{\frac{dk}{2\pi}} \frac{2 \cos n k}{1 - \gamma \cos k} \cos \left( \frac{\pi \ell t}{2 L} \cos \frac{k}{2} \right). \label{nCdef}
\end{align} 
Similarly, we can write down the formal expression for the currents $\mathfrak{j}^{(n)}(x, t)$
\begin{align}
\mathfrak{j}^{(n)}(x, t) = \sum\limits_{\ell=1}^{\infty} \frac{\sin (\pi \ell/2)}{(\pi \ell/2)} \sin \frac{\pi \ell x}{2L} \left[ \displaystyle{\frac{1}{\beta_1}} D^{(n)}_{\ell}(\gamma_1, t) - \displaystyle{\frac{1}{\beta_2}} D^{(n)}_{\ell}(\gamma_2, t) \right], \label{ncurrGGE}
\end{align}
where $D^{(n)}_{\ell}(\gamma, t)$ is given by the integral
\begin{align}
D^{(n)}_{\ell}(\gamma, t) = \int\limits_{0}^{\pi} \displaystyle{\frac{dk}{2\pi}} \frac{2 \cos n k \cos (k/2)}{1 - \gamma \cos k} \sin \left( \frac{\pi \ell t}{2 L} \cos \frac{k}{2} \right). \label{nDdef}
\end{align}
It can be shown that, in the large time limit, the densities and currents in Eq.~\eqref{nrhoGGE} and Eq.~\eqref{ncurrGGE} respectively relax to their equilibrium values as $\sim1/\sqrt{t}$, similar to the finite-size case in Sec.~\eqref{subsec:domainWallGE}. In Figs.~(\ref{2casecgQ0}-\ref{2casecgQ1curr}), we show a comparison of the analytic results from hydrodynamics, for temperature, energy current density, $Q^{(1)}$ density, and $Q^{(1)}$ current density with those obtained numerically from the microscopics by computing averages of the corresponding microscopic quantities in  Eqs.~(\ref{local-Q_n}, \ref{currMicro}) and using the correlations in Eq.~\eqref{qq_pp_mat}. The qualitative features of the evolution of these quantities are similar to those for the finite chain in Sec.~\eqref{subsec:domainWallGE}. We again find a good agreement between hydrodynamics and exact numerics.

\begin{figure}[H]
\hspace*{-1.2 cm}
\centering
	\includegraphics[scale=0.5]{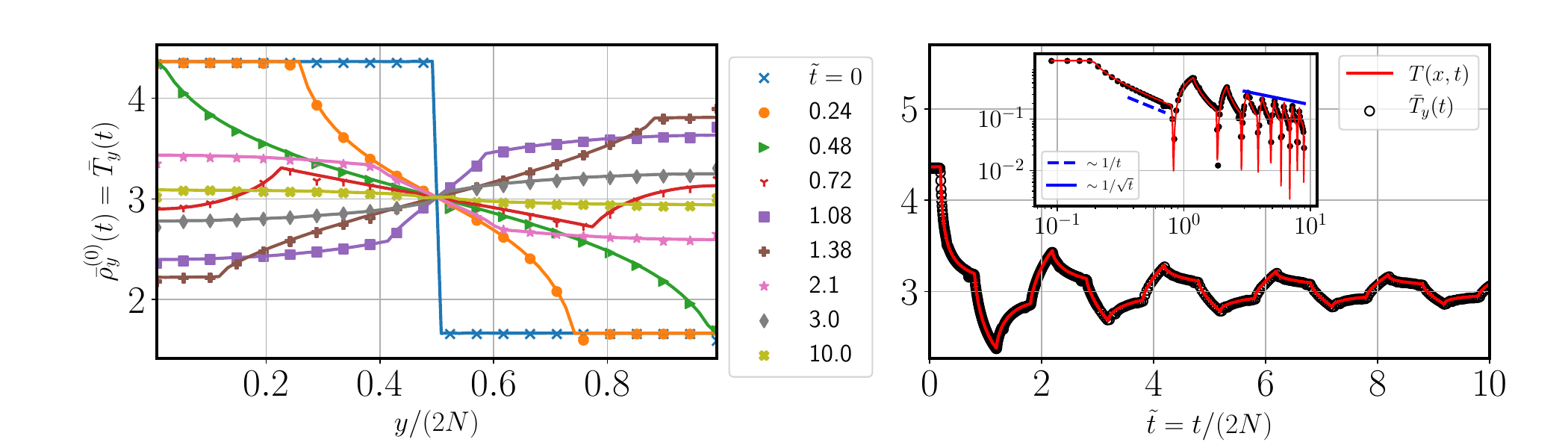}
\begin{tikzpicture}[>=latex]
    \node[anchor=east] at (2,0) {(a)}; 
    \node[anchor=west] at (10.5,0) {(b)};
\end{tikzpicture}
\caption{(a) Plot showing the coarse-grained temperature profile at different times  $\tilde{t}=t/(2N)$. (b) shows the evolution of the temperature with time at a fixed coarse-grained point $y/(2N) \approx 0.30$. The inset in (b) shows the same curve on a log-log scale but with the steady state value subtracted. We observe a $\sim 1/t$ decay at small times which, after many reflections from the boundaries, changes to a $\sim 1/\sqrt{t}$ decay at large times as expected from the analytical expressions. In both figures, the points are calculated by coarse-graining the microscopic temperature profile whereas the continuous lines are obtained from the analytic GHD expressions. We find an excellent agreement between the coarse-grained microscopic and the GHD results. The coarse-graining is done over $\Delta = 8$ consecutive sites. The parameters used are $\beta_1 = 0.25, \beta_2 = 1.0, \gamma_1 = 0.4, \gamma_2 = 0.8$ and $2N=512$.} \label{2casecgQ0}
\end{figure}
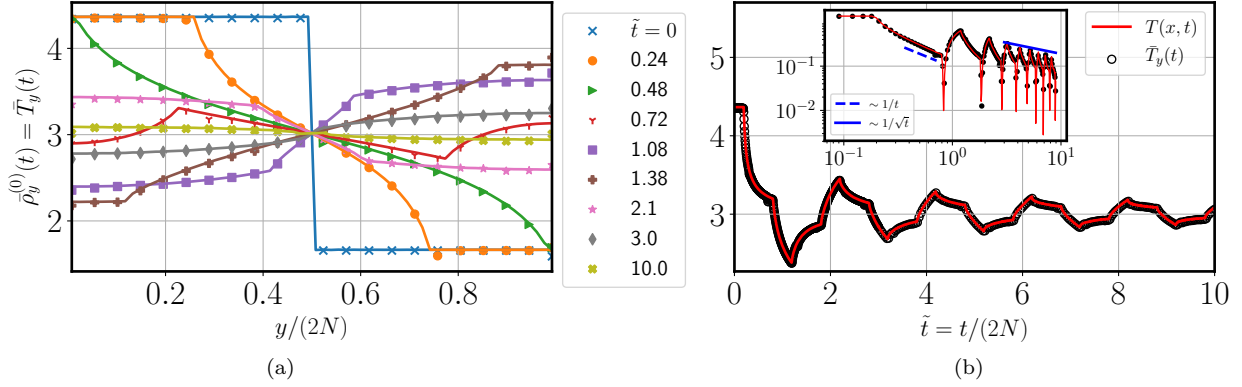

\begin{figure}[H]
\hspace*{-0.5cm}
\centering
	\includegraphics[scale=0.48]{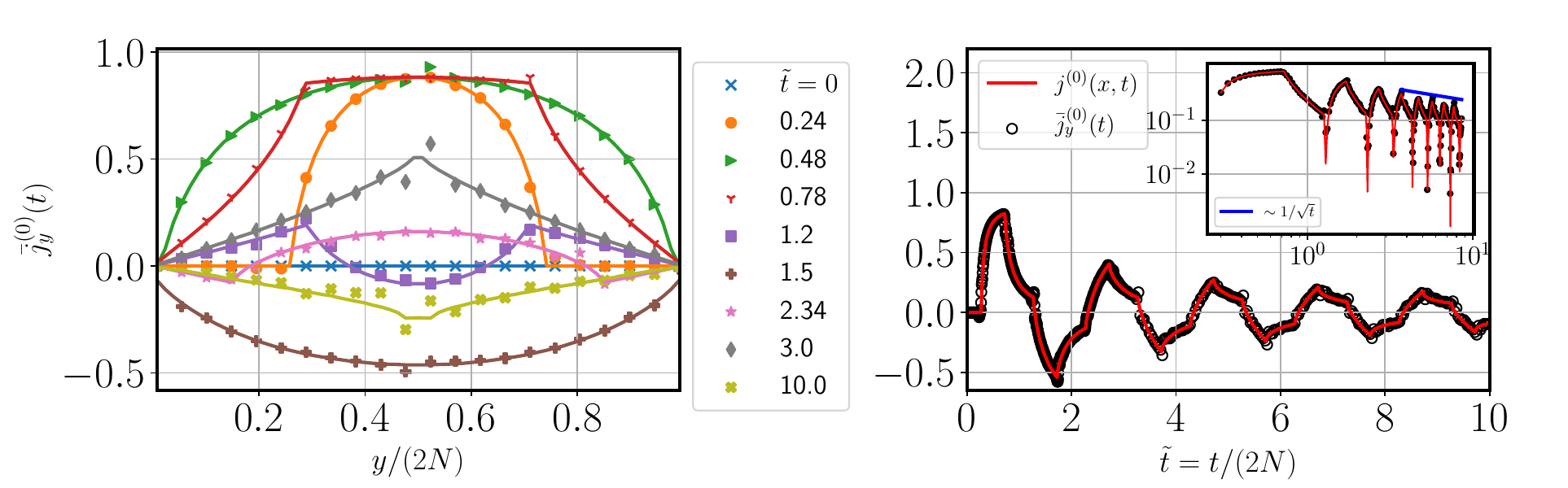}
\begin{tikzpicture}[>=latex]
    \node[anchor=east] at (2,0) {(a)}; 
    \node[anchor=west] at (10,0) {(b)};
\end{tikzpicture}
\caption{(a) Plot showing the coarse-grained energy current density profile at different times  $\tilde{t}=t/(2N)$. (b) shows the evolution of the energy current density with time at a fixed coarse-grained point $y/(2N) \approx 0.23$. The inset in (b) shows the same curve on a log-log scale. We observe a $\sim 1/\sqrt{t}$ decay at large times as expected from the analytical expressions. In both figures, the points are calculated by coarse-graining the microscopic energy current density profile whereas the continuous lines are obtained from the analytic GHD expressions. We find an excellent agreement between the coarse-grained microscopic and the GHD results. The coarse-graining is done over $\Delta = 8$ consecutive sites. The parameters used are $\beta_1 = 0.25, \beta_2 = 1.0, \gamma_1 = 0.4, \gamma_2 = 0.8$ and $2N=512$.} \label{2casecgQ0curr}
\end{figure}

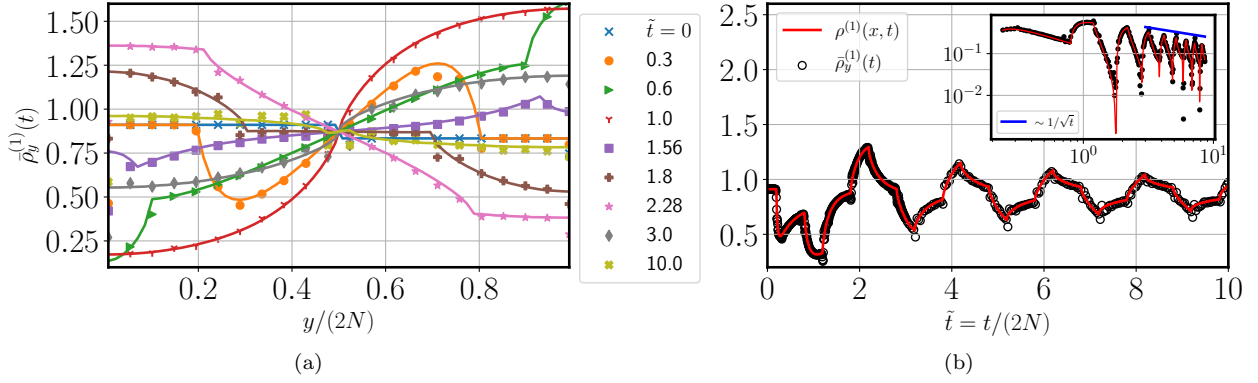
\begin{figure}[H]
\hspace*{-0.8cm}
\centering
	\includegraphics[scale=0.49]{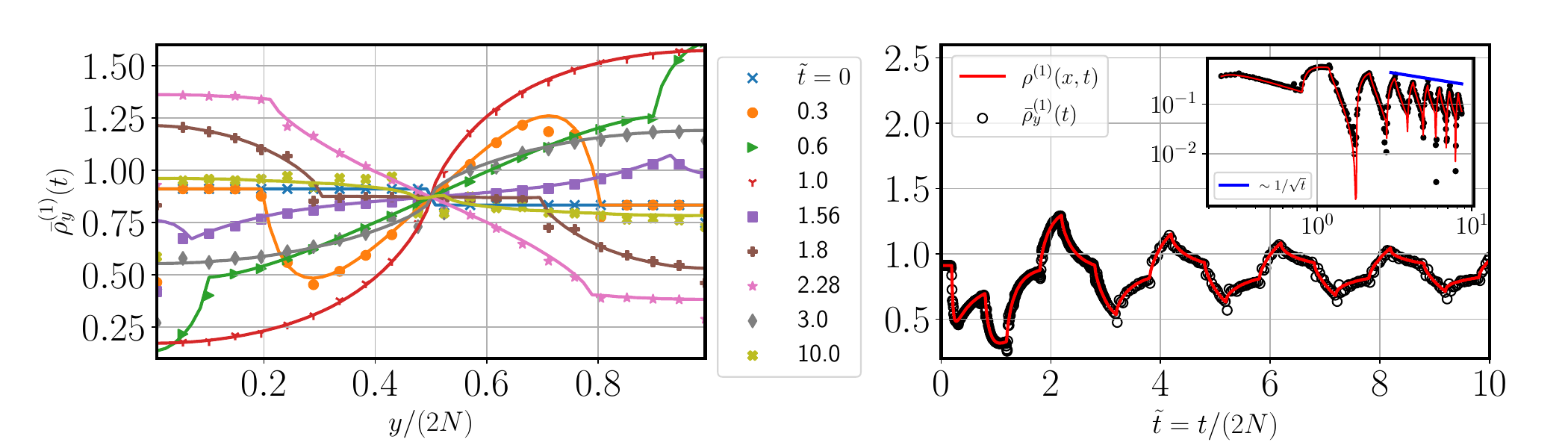}
\begin{tikzpicture}[>=latex]
    \node[anchor=east] at (2,0) {(a)}; 
    \node[anchor=west] at (10,0) {(b)};
\end{tikzpicture}
\caption{(a) Plot showing the coarse-grained $Q^{(1)}$ density profile at different times  $\tilde{t}=t/(2N)$. (b) shows the evolution of the $Q^{(1)}$ density with time at a fixed coarse-grained point $y/(2N) \approx 0.30$. The inset in (b) shows the same curve on a log-log scale but with the steady state value subtracted. We observe a $\sim 1/\sqrt{t}$ decay at large times as expected from the analytical expressions. In both figures, the points are calculated by coarse-graining the microscopic $Q^{(1)}$ density profile whereas the continuous lines are obtained from the analytic GHD expressions. We find an excellent agreement between the coarse-grained microscopic and the GHD results. The coarse-graining is done over $\Delta = 8$ consecutive sites. The parameters used are $\beta_1 = 0.25, \beta_2 = 1.0, \gamma_1 = 0.4, \gamma_2 = 0.8$ and $2N=512$.} \label{2casecgQ1}
\end{figure}

\begin{figure}[H]
\hspace*{-0.4cm}
\centering
	\includegraphics[scale=0.48]{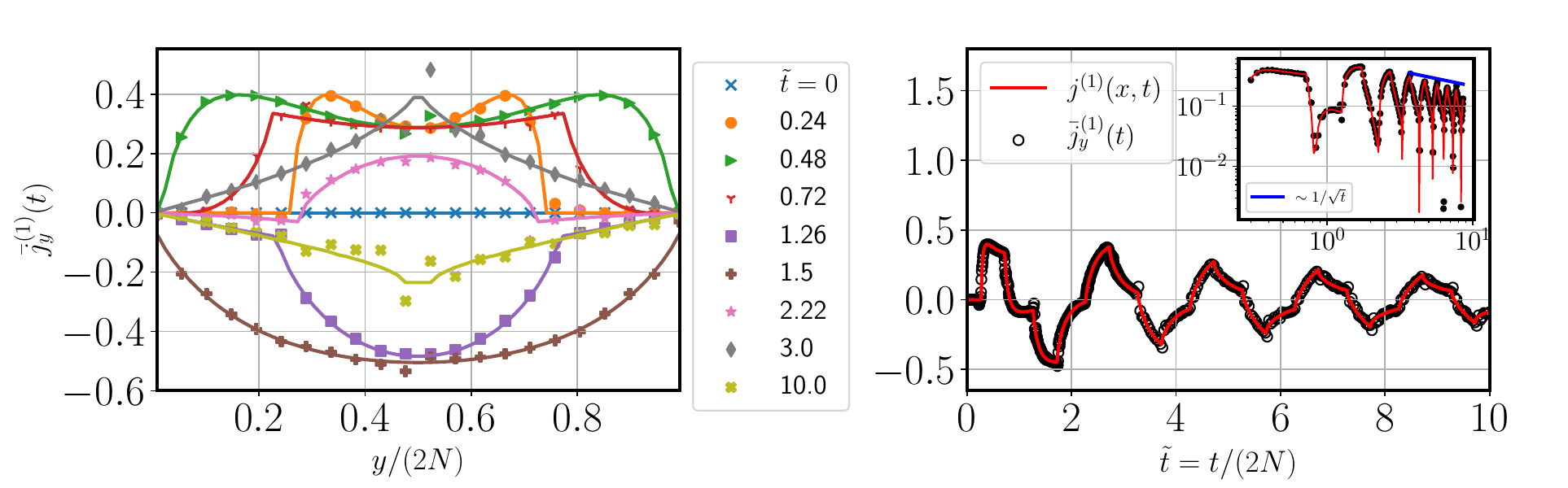}
\begin{tikzpicture}[>=latex]
    \node[anchor=east] at (3,0) {(a)}; 
    \node[anchor=west] at (11.1,0) {(b)};
\end{tikzpicture}
\caption{(a) Plot showing the coarse-grained $Q^{(1)}$ current density profile at different times  $\tilde{t}=t/(2N)$. (b) shows the evolution of the $Q^{(1)}$ current density with time at a fixed coarse-grained point $y/(2N) \approx 0.23$. The inset in (b) shows the same curve on a log-log scale. We observe a $\sim 1/\sqrt{t}$ decay at large times as expected from the analytical expressions. In both figures, the points are calculated by coarse-graining the microscopic $Q^{(1)}$ current density profile whereas the continuous lines are obtained from the analytic GHD expressions. We find an excellent agreement between the coarse-grained microscopic and the GHD results. The coarse-graining is done over $\Delta = 8$ consecutive sites. The parameters used are $\beta_1 = 0.25, \beta_2 = 1.0, \gamma_1 = 0.4, \gamma_2 = 0.8$ and $2N=512$.} \label{2casecgQ1curr}
\end{figure}
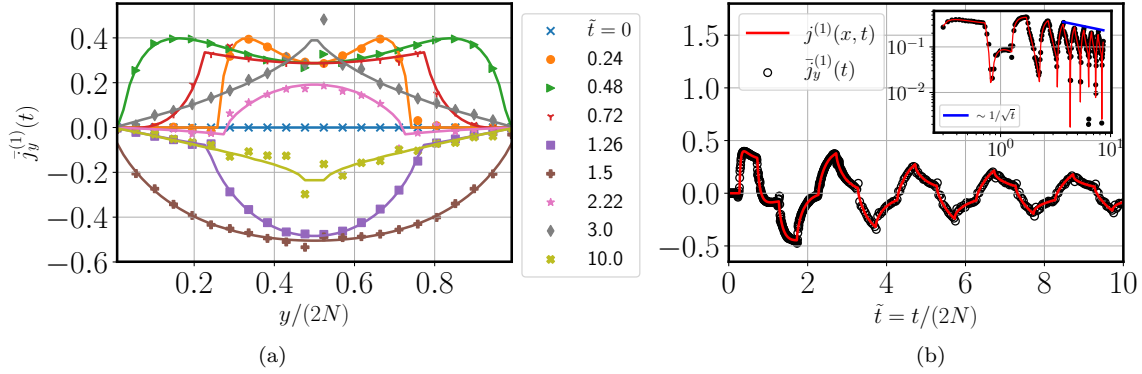

\noindent We expect the final state to be a GGE state characterised by Lagrange multipliers $\{ \lambda_n \}$ which are related to the average densities as
\begin{align}
\rho^{(n)}_{GGE} = \int\limits_{-\pi}^{\pi} \displaystyle{\frac{dk}{2 \pi}} \displaystyle{\frac{\cos nk}{\sum\limits_{m=0}^{\infty} \lambda_m \cos mk}}. \label{rho_lambdaGGE}
\end{align}
The expression in Eq.~\eqref{rho_lambdaGGE} can be easily inverted using the Fourier cosine series and we obtain the following explit expressions for the Lagrange multipliers of the finite chain
\begin{subequations}
\begin{align}
\lambda_0 &= \displaystyle{\frac{1}{2\pi}} \int\limits_{0}^{\pi} dk~\frac{1}{f(k)},  \\
\lambda_n &= \displaystyle{\frac{1}{\pi}} \int\limits_{0}^{\pi} dk~\frac{\cos nk}{f(k)} ~~\text{for} ~~n>1,
\end{align} 
\label{Lagrange_multipliers}
\end{subequations}
where $f(k)$ is given by
\begin{align}
f(k) = \displaystyle{\frac{1}{2}} \rho^{(0)}_{GGE} + \sum\limits_{n=1}^{\infty} \rho^{(n)}_{GGE} \cos nk. \label{fdef}
\end{align}
Putting the values of the conserved densities from Eq.~\eqref{GGEdensity} into Eq.~\eqref{Lagrange_multipliers}, we can compute the $\{ \lambda_n \}$ in the final state. In Fig.~\eqref{rho_lambda}, we plot the stationary state density and the absolute values of the Lagrange multipliers on a log-linear scale. Unlike the case where the initial state is composed of two GEs, here we find that all densities and Lagrange multipliers are non-zero in the final state. However, both quantities decay exponentially for large $n$. Note that Eqs.~(\ref{rho_lambdaGGE}-\ref{fdef}) are valid even if the initial state is described by a product of two GGEs with an arbitrary number of nonzero parameters that have unequal values in the two halves of the chain.

\begin{figure}[H]
\hspace*{-0.4cm}
\centering
	\includegraphics[scale=0.5]{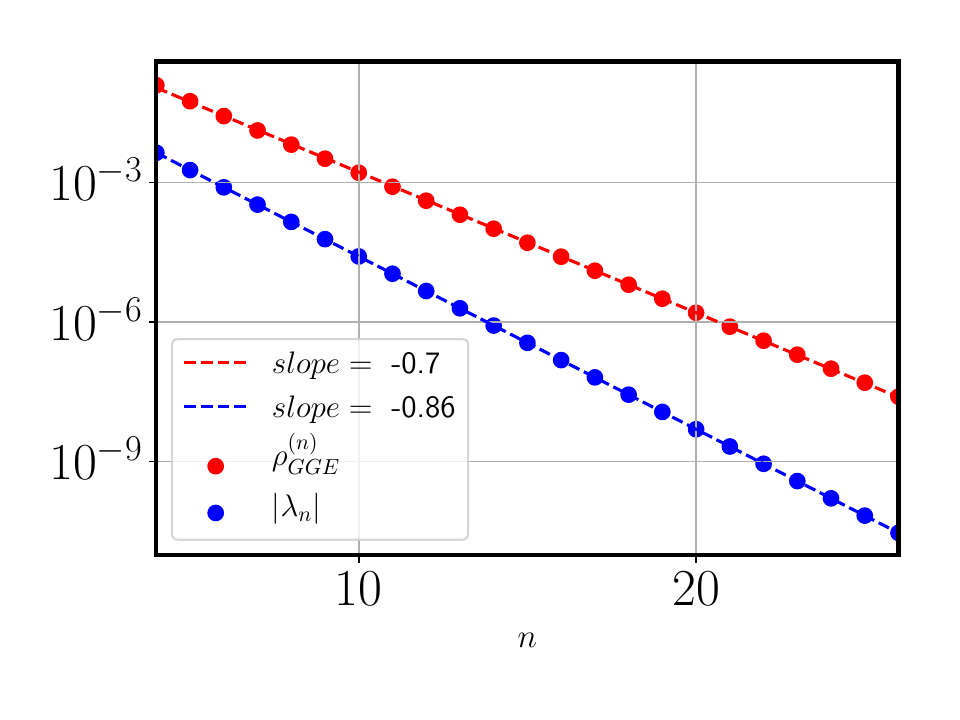}
\caption{Plot showing exponential decay of the densities $\rho^{(n)}_{GGE}$ and the absolute values of the Lagrange multipliers $\lambda_n$ with increasing $n$. The points in red are calculated using Eq.~\eqref{GGEdensity} whereas the blue points are evaluated using Eq.~\eqref{Lagrange_multipliers}. The dashed lines show an exponential fit to the data points. The parameters used are $\beta_1 = 0.25, \beta_2 = 1.0, \gamma_1 = 0.4, \gamma_2 = 0.8$.} \label{rho_lambda}
\end{figure}

\section{Conclusions} \label{sec:conclusions}

In this work, we first provided a physically motivated derivation of the generalized hydrodynamics of a classical harmonic chain, based on the approach of the Wigner function. Our work thus makes connections between earlier work on harmonic chains \cite{dobrushin1986,dobrushin1988,boldrighini1983euler,boldrighini1997navierstokes} with the recent developments in GHD. As an application, we studied the problem of thermalization of a harmonic chain starting from domain wall initial conditions.

{Specifically, we studied the evolution of a classical harmonic chain with nearest-neighbor interactions starting from a domain wall initial state composed of either two GEs or two GGEs with unequal chemical potentials in the left and right halves of the chain. 
For each type of the initial distribution, we study both finite and infinite chains. In all cases, we found excellent agreement, at all times, between results from hydrodynamics with those obtained from exact numerics. 
We observe that a finite chain relaxes as $\sim 1/\sqrt{t}$ to a steady state that is expected to be described by a GE or a GGE depending on whether we use the GE or GGE type of initial condition. For the GGE initial condition, the final state is also expected to be a GGE characterized by a set of Lagrange multipliers which we evaluate exactly in terms of the steady state densities. We find that this GGE steady state has an infinite number of Lagrange parameters with exponentially decaying strengths.
All currents decay to zero as $\sim 1/\sqrt{t}$ as we approach the steady state due to reflections at the boundaries.
For the case of an infinite chain, we find that even for the initial condition composed of two GEs, any finite region of the system tends, in the $t \to \infty$ limit, towards a non-equilibrium steady state (NESS) in which the currents have nonzero but constant values.  The NESS in the infinite chain is therefore expected to be a current-carrying GGE. The relaxation to this NESS happens as $\sim 1/t$ for the densities and as $\sim 1/t^2$ for the currents.}

We note that the Wigner GHD equation is applicable for generic harmonic systems incorporating more complex force matrices in arbitrary dimensions and quantum statistics. These can be interesting extensions of the present work. For the quantum case, the computation of entanglement entropy in free fermionic systems using semiclassical hydrodynamics has generated a lot of interest \cite{bertini2018GHD,saha2024page} and it would be of great interest to explore this in the context of quantum harmonic crystals. Another interesting open question would be to prove or demonstrate equilibration in the harmonic chain for a single realization that starts from an arbitrary initial condition. This has recently been proved~\cite{tasaki2024macroscopic} in the quantum case for non-interacting fermions while for a classical ideal gas, it has been established for initial conditions chosen from continuous distributions~\cite{deBievre2017Kac}.

\section{Acknowledgements} \label{sec:ack}
We thank Herbert Spohn and Benjamin Doyon for very useful discussions.  We acknowledge support from the Department of Atomic Energy, Government of India, under Project No. RTI4001. AD acknowledges the J.C. Bose Fellowship (JCB/2022/000014) of the Science and Engineering Research Board of the Department of Science and Technology, Government of India. A. K. would like to acknowledge the support of DST, Government of India Grant under Project No. CRG/2021/002455  and the MATRICS grant MTR/2021/000350 from the SERB, DST, Government of India.

\appendix

\section{GGE partition function} \label{app:Z_GGE}
The GGE partition function in Eq.~\eqref{P_GGE} is given by the integral
\begin{align}
\begin{split}
Z(\{ \lambda_n \}) &:= \int_{-\infty}^{\infty} \prod_{j=1}^N dp_j dq_j \exp \left( -\sum\limits_{n=0}^{N-1} \lambda_n Q^{(n)}({\bf p}, {\bf q}) \right) \\
&= \int_{-\infty}^{\infty} \prod_{j=1}^N dp_j dq_j \exp \left( -\frac{1}{2} \left[ {\bf{p}}^T \sum_n \lambda_n B^{(n)} {\bf{p}} 
 + {\bf{q}}^T \sum_n \lambda_n M^{(n)} {\bf{q}} \right] \right)
\end{split} \label{Z_GGE1}
\end{align}
where we have used the expression for $Q^{(n)}$ in Eq.~\eqref{Qnmat} and the Lagrange multipliers are chosen such that $\sum_n\lambda_n B_n$ and $\sum_n\lambda_n M_n$ are positive definite matrices. The Gaussian integration can be easily performed once we know the eigenvalues of $B^{(n)}$ and $M^{(n)}$ matrices which are respectively given by $\cos (n k_{\ell})$ and $\omega_{k_{\ell}}^2 \cos (n k_{\ell})$ with $k_{\ell} = \displaystyle{\frac{\pi \ell}{N+1}}, ~\ell = 1, \ldots, N$ and $\omega_{k}^2 = 2(1-\cos k)$.
We therefore get
\begin{align}
Z(\{ \lambda_n \}) = \sqrt{\displaystyle{\frac{(2\pi)^N}{\prod_{\ell=1}^N \left( \sum_{n=0}^{N-1} \lambda_n \cos (n k_{\ell}) \right)}}} \times \sqrt{\displaystyle{\frac{(2\pi)^N}{\prod_{\ell=1}^N \omega_{k_{\ell}}^2 \left( \sum_{n=0}^{N-1} \lambda_n \cos (n k_{\ell}) \right)}}}, \label{Z_GGE2}
\end{align}
which can be simplified to yield Eq.~\eqref{partitionFunc}.

\section{Density evolution corresponding to the initial condition composed of two GGEs} \label{app:GGE}
Here we show the steps leading to the density profiles in Eq.~\eqref{nrhoGGEInfinite2} starting from the expression for the Wigner function in Eq.~\eqref{wignerInfiniteGGE1}. Substituting the Wigner function from Eq.~\eqref{wignerInfiniteGGE1} into Eq.~\eqref{nrho}, we obtain the following expression for $\rho^{(n)}(x, t)$
\begin{align}
\rho^{(n)}(x, t) = 2 \int\limits_0^{\pi/2} \displaystyle{\frac{d \theta}{\pi}} \cos 2 n \theta \frac{1}{\beta_1(1 - \gamma_1 \cos 2\theta)} - I^{(n)}(x, t) = g^{(n)}(\beta_1, \gamma_1) - I^{(n)}(x, t), \label{nrhoGGEInfinite1}
\end{align}
where $g^{(n)}(\beta, \gamma)$ is defined in Eq.~\eqref{ngdef} and $I^{(n)}(x, t)$ is given by
\begin{align}
I^{(n)}(x, t) = \int\limits_0^{\pi/2} \displaystyle{\frac{d \theta}{\pi}} \cos 2 n \theta \left( \frac{1}{\beta_1(1 - \gamma_1 \cos 2 \theta)} - \frac{1}{\beta_2(1 - \gamma_2 \cos 2 \theta)} \right) \left[ \Theta(x - t \cos \theta) + \Theta(x + t \cos \theta) \right]. \label{IGGE1}
\end{align}
Differentiating Eq.~\eqref{IGGE1} w.r.t $x$ on both sides and simplifying yields
\begin{align}
\displaystyle{\frac{\partial I^{(n)}}{\partial x}} = \displaystyle{\frac{\cos 2n \theta^{\star}}{\pi t \sin \theta^{\star}}}\left( \frac{1}{\beta_1(1 - \gamma_1 \cos 2 \theta^{\star})} - \frac{1}{\beta_2(1 - \gamma_2 \cos 2 \theta^{\star})} \right) , ~|x| < t \label{IGGE2}
\end{align}
where $\theta^{\star} = \cos^{-1}(|x|/t)$. Integrating Eq.~\eqref{IGGE2} w.r.t $x$ gives
\begin{align}
I^{(n)}(x, t) = 
\begin{cases}
0, ~ x < -t \\ \\
\bigintss\limits_0^{\pi/2} \displaystyle{\frac{d \theta}{\pi}} \cos 2 n \theta \left( \frac{1}{\beta_1(1 - \gamma_1 \cos 2 \theta)} - \frac{1}{\beta_2(1 - \gamma_2 \cos 2 \theta)} \right) \\
+ \bigintss\limits_{0}^{\sin^{-1}(x/t)} \displaystyle{\frac{d\theta}{\pi}} \cos \{ 2 n \cos^{-1} |\sin \theta| \} \left( \frac{1}{\beta_1(1 + \gamma_1 \cos 2\theta)} - \frac{1}{\beta_2(1 + \gamma_2 \cos 2\theta)} \right), ~ |x| < t \\ \\
2 \bigintss\limits_0^{\pi/2} \displaystyle{\frac{d \theta}{\pi}} \cos 2 n \theta \left[ \left( \frac{1}{\beta_1(1 - \gamma_1 \cos 2 \theta)} - \frac{1}{\beta_2(1 - \gamma_2 \cos 2 \theta)} \right) \right], ~ x > t
\end{cases} \label{IGGE3}
\end{align}
Using Eq.~\eqref{ngdef}, we can rewrite Eq.~\eqref{IGGE3} as
\begin{align}
I^{(n)}(x, t) = 
\begin{cases}
0, ~ x < -t \\ \\
\displaystyle{\frac{1}{2}} [g^{(n)}(\beta_1, \gamma_1) - g^{(n)}(\beta_2, \gamma_2)] 
\\ 
+ \bigintss\limits_{0}^{\sin^{-1}(x/t)} \displaystyle{\frac{d\theta}{\pi}} \cos \{ 2 n \cos^{-1} |\sin \theta| \} \left( \frac{1}{\beta_1(1 + \gamma_1 \cos 2\theta)} - \frac{1}{\beta_2(1 + \gamma_2 \cos 2\theta)} \right), ~ |x| < t \\ \\
g^{(n)}(\beta_1, \gamma_1) - g^{(n)}(\beta_2, \gamma_2), ~ x > t
\end{cases} \label{IGGE4}
\end{align}
Substituting Eq.~\eqref{IGGE4} into Eq.~\eqref{nrhoGGEInfinite1} gives us the required expression in Eq.~\eqref{nrhoGGEInfinite2}.

}

\bibliographystyle{ieeetr}
\bibliography{bibliography.bib}

\end{document}